\documentclass[
 reprint,
 groupedaddress,
 showpacs,
 superscriptaddress, 
 amsmath,amssymb,
 aps,
 pre,
 nofootinbib,
 twocolumn
]{revtex4-1}

\usepackage{amssymb}
\usepackage{amsmath}
\usepackage{physics}
\usepackage{empheq}
\usepackage{amsthm}
\usepackage{mathtools}
\usepackage[pdftex]{graphicx}
\usepackage[short]{turnthepage}
\usepackage{color}
 \usepackage[driverfallback=dvipdfmx]{hyperref}
 \usepackage{url}
 
\hypersetup{colorlinks=true
  ,urlcolor=blue
  ,anchorcolor=blue
  ,citecolor=magenta
  ,filecolor=blue
  ,linkcolor=blue
  ,menucolor=blue
  ,linktocpage=true
  ,pdfproducer=medialab}

\DeclareMathOperator{\sgn}{sgn}
\mathtoolsset{showonlyrefs=true}

\theoremstyle{definition}
\newtheorem{theorem}{Theorem}

\newcommand{\ifrac}[2]{{\left. #1 \middle/ #2 \right. }}
\newcommand{\siki}[1]{Eq.~\eqref{#1}}
\newcommand{\zu}[1]{Fig.\ref{#1}}
\newcommand{\Sec}[1]{Sec.~\ref{#1}}
\newcommand{\sr}[1]{Sec.~{#1}}

\newcommand{\app}[1]{Appendix \ref{#1}}

\newcommand{\sorder}[1]{o\qty(#1)}

\newcommand{\kak}[1]{\biggl( #1 \biggr)}
\newcommand{\Np}{{N_\mathrm{para}}}
\newcommand{\Nt}{{N_\mathrm{tune}}}
\newcommand{\Nf}{{N_\mathrm{free}}}

\newcommand{\ta}{\theta}
\newcommand{\za}{\zeta}

\makeatletter
\renewcommand{\p@subsection}{\thesection.}
\renewcommand{\p@subsubsection}{\thesection.\thesubsection.}
\makeatother

\allowdisplaybreaks

\begin{document}
\begin{flushright}
KUNS-2906
\end{flushright}

\title{Path integral approach to universal dynamics of reservoir computers}
\author{Junichi Haruna}
\email{j.haruna@gauge.scphys.kyoto-u.ac.jp}
% \thanks{These two authors contributed equally }
\affiliation{%
 Department of Physics, Kyoto University, Kyoto 606-8502, Japan
}%

\author{Riki Toshio}
\email{toshio.riki.63c@st.kyoto-u.ac.jp}
% \thanks{These two authors contributed equally }
\affiliation{%
 Department of Physics, Kyoto University, Kyoto 606-8502, Japan
}%

\author{Naoto Nakano}
\email[]{n\_nakano@meiji.ac.jp}
\affiliation{%
Graduate School of Advanced Mathematical Sciences, Meiji University, Tokyo 164-8525, Japan
}%

\begin{abstract}
In this work, we give a characterization of the reservoir computer (RC) by the network structure, especially the probability distribution of random coupling constants.
First, based on the path integral method, we clarify the universal behavior of the random network dynamics in the thermodynamic limit, which depends only on the asymptotic behavior of the second cumulant generating functions of the network coupling constants.
This result enables us to classify the random networks into several universality classes, according to the distribution function of coupling constants chosen for the networks.
Interestingly, it is revealed that such a classification has a close relationship with the distribution of eigenvalues of the random coupling matrix.
We also comment on the relation between our theory and some practical choices of random connectivity in the RC.
Subsequently, we investigate the relationship between the RC's computational power and the network parameters for several universality classes.
We perform several numerical simulations to evaluate the phase diagrams of the steady reservoir states, common-signal-induced synchronization, and the computational power in the chaotic time series inference tasks. 
As a result, we clarify the close relationship between these quantities, especially a remarkable computational performance near the phase transitions, which is realized even near a non-chaotic transition boundary. 
These results may provide us with a new perspective on the designing principle for the RC.

\end{abstract}
% \preprint{KUNS-00000}

\date{\today}

\pacs{87.19.lj, 64.60.aq, 84.35.+i, 05.45.Tp, 87.10.-e}

\maketitle

\section{Introduction}
Artificial neural networks, originally designed to emulate biological network systems such as the human brain~\cite{Rabinovich2006}, serve as an essential basis of modern machine learning theory.
These systems, especially recurrent random network models, exhibit rich dynamical properties, including collective chaotic dynamics~\cite{Sompolinsky1988, Vreeswijk1996, Toyoizumi2011, Kadmon2015, Keup2021}, common-signal-induced synchronization~\cite{Toral2001, Jaeger2001, Zhou2002, Teramae2004, Uchida2004,Lu2018}, and noise-induced suppression of chaos~\cite{Molgedey1992, Rajan2010, Schuechker2018}.
These dynamical aspects of random networks were investigated in great detail in early seminar works~\cite{Amari1972,Parisi1986,Sompolinsky1988,Molgedey1992,Vreeswijk1996} and, more recently, formulated elegantly with the generating functional formalism~\cite{Schuechker2018, Crisanti2018}.

Besides, it is pretty remarkable that common signal-induced synchronization, where any trajectory of the reservoir state converges under the same input regardless of the previous history, brings a new informational perspective to the network dynamics. 
Accompanied by the rich dynamical patterns of the networks, this phenomenon enables them to serve as a resource for real-time information processing, such as time series prediction~\cite{Jaeger2004,Li2012,Pathak2017,Pathak2018}, observer problem~\cite{Lu2017,Zimmermann2018}, reinforcement learning~\cite{Bush2005,Legenstein2008}, and speech recognition~\cite{Verstraeten2005,Verstraeten2007,Skowronski2006}.
Such a framework is referred to as {\it reservoir computing} (RC) (for reviews, see \cite{Lukosevicius2009,Lukosevicius2012,Tanaka2019}), which was originally proposed in the context of machine learning~\cite{Jaeger2001, Jaeger2004} (called Echo state networks: ESNs) and computational neuroscience~\cite{Maass2002}.
Furthermore, there is also a new ambitious attempt to exploit the complex dynamics of real physical systems as information-processing devices based on the RC scheme, which is referred to as {\it physical reservoir computing} (for reviews, see \cite{Tanaka2019,Nakajima2020}).

In the last decades, the RC has been generalized to various network architectures, such as the backpropagation decorrelation~\cite{Steil2004} and the FORCE learning~\cite{Sussillo2009}, to improve the computational performance or to clarify the working principles of the RC~\cite{Xue2007, Schrauwen2008,Rodon2011,Cui2012, Laje2013,Farkas2016, Gallicchio2017,Malik2017, Inubushi2017, Carroll2019,Tanaka2020}.
Their information processing capacities have ever been evaluated with various informational or dynamical measures~\cite{Ozturk2007,Verstraeten2007,Ganguli2008, Dambre2012, Carroll2020a}.
However, despite these efforts for many years, we have yet understood enough how to design or tune the reservoir network to optimize its computational performance.

The {\it Edge of chaos}, a concept initially introduced in the context of a cellular automaton~\cite{Langton1990}, is one of the most common guiding principles to tune the parameters of RCs.
Reservoir networks are known to exhibit various dynamical phases, such as chaotic or ordered phases, depending on the values of the parameters.
Although some controversies exist~\cite{Carroll2020b}, RCs are believed to show a long temporal memory or the best computational performance near the boundary between ordered and chaotic phases, that is, the edge of chaos~\cite{Bertschinger2004,Bertschinger2005, Legenstein2007a,Legenstein2007b,Busing2010, Boedecker2012, Haruna2019}.  
Additionally, it has also been shown that decoding signals from these networks are robust to noise above and near the transition to chaos~\cite{Toyoizumi2011}.
These dynamical features of neural networks have ever been attracting much attention even in neuroscience because of several pieces of evidence that the cortical circuit has an architecture tuned to such a critical state~\cite{Beggs2003,Chialvo2004,Beggs2008}.
These observations suggest that it is crucial to the design of the RC to clarify the close relationship between the informational processing ability of the RC and the dynamical properties of the random network.

In this work, we aim to characterize the RC by the network structure and propose a clue to better RC's design.
For this purpose, we analyze the dependence of the network dynamics on the probability distribution from which the coupling constants are sampled, based on theoretical and numerical approaches as follows.

First, using the generating functional formalism for random neural networks, we clarify the universality of the network dynamics in the large-$N$ limit, where the network size $N$ becomes infinitely large. 
We show that the reservoir networks with any probability distribution of the coupling constant are classified into several universality classes according to the asymptotic behavior of their second generating functions in the limit.
The following numerical simulations demonstrate that networks in an identical class yield equivalent dynamics.
Particularly important is that, for some universality classes, higher-order statistics play a crucial role in determining the network dynamics and its computational performance.
This contrasts with the case of the Gaussian networks, which have been most studied so far and can be described only by the one- and two-point correlation functions of the reservoir states~\cite{Sompolinsky1988,Schuechker2018,Crisanti2018}.
Interestingly, we can prove that each universality class has a one-to-one correspondence with the eigenvalue spectrum of the random coupling matrix.
Furthermore, we give some instructive comments on the relation between our theory and practical network structures in implementing RCs.
They include discussions on the sparsity of coupling constants and their normalization with the spectral radius.

Next, to verify the validity of our analytical expectation and the relation with the edge of chaos, we provide several numerical simulations of random network dynamics without and with driving input series.
In the autonomous case, we numerically analyze the asymptotic steady state (or the states after reaching the attractor) of the networks and show the phase diagrams in the parameter space for each universality class.
We also estimate the finite-size effects in their Lyapunov exponent. 
Then, we consider common-signal-induced synchronization and a time series inference task of chaotic signals in input-driven networks.
Jaeger \cite{Jaeger2002} showed that common-signal-induced synchronization realizes when the system exhibits the echo-state property. Thus we investigate which parameters in the phase space achieve the synchronization.
Moreover, we attempt to perform time series forecasting task using the RC with output feedback. For this task we calculate its maximum Lyapunov exponent to verify the performance of dynamical reconstruction.
More specifically, we assume the time series of the $x$-coordinate of the Lorenz system~\cite{Lorenz1963} as input time series and attempt to infer the concurrent values of other coordinates~\cite{Lu2017}.

The synchronization simulation suggests that the chaotic input signals shrink an area of the chaotic phase.
This result is consistent with the one discussed in previous studies~\cite{Molgedey1992,Rajan2010,Schuechker2018}.
In the time series inference tasks with an open-loop or a closed-loop system, we observe a remarkable improvement in the RC's performance at the parameters with which the system undergoes phase transitions.
Although our result for the Gaussian network is essentially equivalent to the one discussed in previous studies~\cite{Bertschinger2004,Bertschinger2005, Legenstein2007a,Legenstein2007b,Busing2010, Boedecker2012, Haruna2019}, it provides us with a new perspective on the boundary between chaotic and polarized ordered phases.
Particularly interesting is that the Gamma network, where higher-order statistics play a crucial role, exhibits computational improvement even near the boundary of a non-chaotic phase. 
This result may contrast with our conventional understanding of the edge of chaos.

This paper is organized as follows.
In \sr{\ref{s:SummaryTh}}, we give a brief summary of the theoretical parts (\sr{\ref{s:RevPI}} - \sr{\ref{s:Generalization}}).
\Sec{s:RevPI} reviews the generating function formalism for random neural networks with path integral representations.
We discuss the familiar dynamical mean-field theory for the averaged network dynamics in the large-$N$ limit and generalize it to more general problems.
In \Sec{s:Universality}, we calculate the concrete form of $K(q)$ defined in \siki{e:DefK} for some distributions of $P_J$.
The result can be classified into several ``universality" classes, and we give some examples.
We also argue the relationship between each universality class and the eigenvalue spectrum of the random coupling matrix $J_{ij}$.
In \Sec{s:Generalization}, we consider a practical choice of the network structures, such as not fully connected networks or controlling the spectral radius of $J_{ij}$, and argue their universality.
In \Sec{s:NumCalc}, we give numerical simulations of random network dynamics without and with driving input signals.
We study the phase diagrams in the parameter space and the Lyapunov exponent in the former case.
In the latter case, we consider common signal-induced synchronization and a time series inference task of chaotic signals.
Additionally, We attempt to construct a simulator of chaotic time series without driving input signals based on an output feedback loop system.
\Sec{s:Conclusion} is devoted to discussion and conclusion.

\section{Summary of theoretical part}
\label{s:SummaryTh}

In this section, we give a summary of our theoretical results, especially for readers who are unfamiliar with the path-integral formalism and seek to utilize our theoretical results for actual numerical simulations.
If the readers want to skip the theoretical details (Sec.~\ref{s:RevPI} and Sec.~\ref{s:Universality}), they can proceed to Sec.~\ref{s:NumCalc} after reading this section.

In this paper, we study a recurrent random network model with nodes $r_i(t)\in \mathbb{R}\,(i=1,\ldots,N)$ that obey the differential equation in continuous time $t$,
\begin{equation}
\label{network eq}
    \dv{r_i}{t} {(t)} = -r_i(t) + \sum_{j=1}^N J_{ij} \phi(r_j(t)) + b_i(t),
\end{equation}
with an initial condition of $r_i(t)=r_i^0$.
$b_i(t)$ is an external input.
$\phi(x)$ is the activation function.
While the activation function is usually taken as a sigmoid-type function $\phi(x)=\mathrm{tanh}(x)$~\cite{Jaeger2001, Lu2017, Sompolinsky1988, Molgedey1992, Rajan2010, Stern2014, Aljadeff2015, Crisanti2018, Schuechker2018} or the Heaviside function $\phi(x) = 1 \ (x \ge 0), \ 0 \ (x < 0)
$~\cite{Bertschinger2004, Bertschinger2005, Kusmierz2020},
we take it as an arbitrary function for generality in what follows.
$J_{ij}$ is a coupling constant between nodes and sampled from an independent and identical distribution $P_J(J_{ij};\theta_a(N))$, which has $N$-dependent tuning parameters $\{\theta_a(N)\}$ in general.
Although the function $P_J(J_{ij};\theta_a(N))$ is usually selected as a Gaussian variate with variance $\sigma^2/N$ and zero mean in the formalism of the path integral method~\cite{Crisanti2018, Schuechker2018}, we consider an arbitrary distribution in this work.

In Sec.~\ref{s:RevPI}, we consider the $J_{ij}$-averaged network dynamics in the thermodynamic limit, where the number of nodes $N$ becomes infinitely large $N\to \infty$.
To summarize our results, we show that the network dynamics in Eq.~\eqref{network eq} is 
asymptotically described by the effective equation of motion
\begin{equation}
\label{e:summary eq}
    \dv{r_i}{t} {(t)} = -r_i(t)  + \eta(t)+ b_i(t).
\end{equation}
Here $\eta(t)$ denotes an effective random noise and its time-correlation functions are given by
\begin{equation}
    \ev{[\eta(t_1)\cdots\eta(t_n)]_C}_\eta=\kappa_nC^*_n(t_1,\cdots,t_n),
\end{equation}
where $[\cdots]_C$ denotes the cumulant of random variables $\eta(t)$, and the coefficient $\kappa_n$ is defined as $\kappa_n\coloneqq i^{-n} d^nK(q)/dq^n\eval{}_{q=0}$ using the 
asymptotic function,
\begin{align}
     K(q) \coloneqq \lim_{N\to\infty} NK_J(q;\theta_a(N)).
\end{align}
Here $K_J(q)$ is the second cumulant generating function of the probability distribution of $J_{ij}$ (See also the definition in \siki{e:DefKJ}).
$C_n^*(t_1,\cdots,t_n)$ are the $n$-point connected correlation functions of $\phi(r_i(t))$ which are averaged over $J_{ij}$-ensembles in the sense of the dynamical mean field approximation. For example, assuming that $J_{ij}$ is sampled from the Gaussian distribution with zero mean and finite variance $\bar{\sigma}/\sqrt{N}$, we obtain $K(q)=-\bar{\sigma}^2q^2/2$, leading to $\kappa_2 = \bar{\sigma}^2$ and $\kappa_i=0$ ($i\neq 2$). 
Importantly, Eq.~\eqref{e:summary eq} means that, in the large-$N$ limit, the coupling terms between nodes in Eq.~\eqref{network eq} are effectively replaced by the random time series $\eta(t)$, which describes an effective force resulting from $J_{ij}$-averaged dynamics of the network nodes. 

From the above result, we notice that effective network dynamics in Eq.~\eqref{e:summary eq} is characterized only by the form of the function $K(q)$ in the thermodynamic limit.
We devote \Sec{s:Universality} to analyzing the forms of $K(q)$ for various choices of probability distributions $P_J(J_{ij})$ and how to tune their parameters in the large-$N$ limit. 
Then, we find that some classes of typical probability distributions lead to the same form of $K(q)$, when their parameters are appropriately tuned for $K(q)$ to be finite in the large-$N$ limit.
We classify typical probability distributions into several classes according to the form of $K(q)$ and call them "universality classes", which includes the {\it Delta class}, the {\it Gauss class}, the {\it stable class}, and the {\it Gamma class}. 
For convenience, we have shown the typical examples in Table.~\ref{t:UnivClass}, which includes the form of $P_J(J_{ij})$, the corresponding form of $K_J(q)$ and $K(q)$, and the way to tune the parameter to make $K(q)$ finite.

More interestingly, we can give a more intuitive meaning to the classification of universality classes.
In the end of \Sec{s:Universality}, we show that each universality class, i.e., each form of $K(q)$, has a one-to-one corresponds with the eigenvalue spectrum of the random matrix $J_{ij}$ in the $N\to\infty$ limit. This means that, in the limit, the information of the eigenvectors of $J_{ij}$ vanishes and the network dynamics is determined only by the eigenvalue spectrum of $J_{ij}$.
This suggest that dynamical properties of the random network, such as dynamical stability and spectral properties of linear response, crucially differ
depending on the universality class, as discussed in Ref.~\cite{Rajan2006,Ahmadian2015}.

As discussed in Sec.~\ref{s:Generalization} in detail, these analyses can be extended to more practical cases where the network connections are not fully connected but sparse, or where the coupling constants $J_{ij}$ are tuned by rescaling the spectral radius. Interestingly, we have shown that, in the latter case, the network dynamics always fall into the Gauss class for any choice of probability distributions. 
This point will be quite important to seek novel types of reservoir computers beyond previous studies.

\section{Path Integral Representation}
\label{s:RevPI}
This section briefly reviews the generating functional formalism for 
random networks in the path integral representation~\cite{Sompolinsky1982, Crisanti1987, Crisanti2018, Schuechker2018}.
We consider the thermodynamic limit, where the number of nodes $N$ becomes infinitely large $N\to \infty$.
We derive the Dynamical Mean-Field (DMF) equation, a technique initially developed for spin glasses~\cite{Sompolinsky1981, Sompolinsky1982} and applied it to describe the time evolution of the averaged network dynamics in the thermodynamic limit~\cite{Sompolinsky1988, Molgedey1992, Rajan2010,Toyoizumi2011, Kadmon2015}.

In the following, we attempt to generalize these well-known frameworks to more general random neural networks with an arbitrary activation function and random neural connectivity.
We discuss the universality in the dynamical properties in \Sec{s:Universality}.

\subsection{Generating Functional Formalism in Path Integral Representation}
\label{S:generating fn.}

We are interested in the statistical aspects of the solutions of Eq.~(\ref{network eq}) (denoted as $r_i^*(t)$) and their correlation functions, such as
$\ev{r_{i_1}^*(t_1)\cdots r_{i_n}^*(t_n)}_J$. 
$r_i^*(t)$ depends on the realization of $J_{ij}$ and $\ev{\cdots}_J$ means the average over $P_J$.
To this end, it is convenient to consider the generating functional for $r_i(t)$ with the Martin–Siggia–Rose–de Dominicis–Janssen (MSRDJ) path integral formalism~\cite{Martin1973, Dominicis1978, Janssen1976, Altland,Taeuber}, which has been used so far to describe, for example, dynamic critical phenomena~\cite{Dominicis1978, Janssen1976, Hohenberg1977} and the dynamics of spin glass systems~\cite{Sompolinsky1982, Crisanti1987} in the context of condensed matter physics.
In our case, the generating functional is defined as the moment-generating function of $r_i(t)$, whose $J$-averaged distribution is given by 
\begin{align}
P_r[r_i(t)]=\ev{\Pi_{i,t}\delta(r_i(t)-r_i^*(t))}_J.
\label{e:prob_r}
\end{align}
The MSRDJ generating function is defined as the characteristic function of $P_r[r_i(t)]$ with an argument $\hat{b}_i(t)$.
Introducing the path integral form with an auxiliary field $\hat{r}_i(t)$, we can represent the generating function as (for details, see the Ref.~\cite{Crisanti2018,Schuechker2018})
\begin{multline}
\label{generating fn.}
    Z[b,\hat{b}] = \int D\hat{r}Dr 
    \int [dJ_{ij}]
    \\
    \prod_{ij} P_J(J_{ij}) 
    \exp(iS + \int dt\sum_{i=1}^N \hat{b}_ir_i),
\end{multline}
where $S[r,\hat{r}]$ is the action of $\hat{r}_i$ and $r_i$ given by
\begin{multline}
    S[r,\hat{r}] \coloneqq
    \\
     \int_0^\infty dt \sum_{i=1}^N \hat{r}_i \kak{-\dv{t} r_i
    -r_i + \sum_{j=1}^N J_{ij} \phi(r_j) + b_i
    }.
    \label{e:DefS}
\end{multline}
Here $[dJ_{ij}]\coloneqq \prod_{i,j=1}^N dJ_{ij}$ denotes the integral measure of $J_{ij}$. 
$Dr$ is defined as $Dr \coloneqq \prod_{i=1}^N Dr_i$ and each $Dr_i$ represents the summation over any paths with the initial condition $r_i(t)=r_i^0$.
For the normalization condition, the generating functional always satisfies the identity $Z[b,0]=1$.

Once the generating functional $Z[b,\hat{b}]$ is obtained, we can calculate the correlation functions or response functions of the states $r_i(t)$ by calculating the functional derivative
\begin{align}
\label{e:correlation_function}
    \ev{ r_{i_1}(t_{i_1})\cdots r_{i_n}(t_{i_n})
    }=
    \fdv{\hat{b}_{i_1}(t_{i_1})} \cdots \fdv{\hat{b}_{i_n}(t_{i_n})}
     Z\eval{}_{\hat{b}_i=0},
\end{align}
where $\ev{\cdots}$ is the average over the paths weighted by the factor $e^{iS}$ and ${\delta}/{\delta\hat{b}_i}$ denotes the functional derivative with respect to $\hat{b}_i(t)$. 
Note that since $P_r[r_i(t)]\neq 0$ only if $r_i(t) = r_i^*(t)$ for each time $t$, correlation functions of $r^*_i$ are equivalent to ones with $r_i$:
\begin{align}
    \ev{ r^*_{i_1}(t_{i_1})\cdots r^*_{i_n}(t_{i_n})}_J
    =
    \ev{ r_{i_1}(t_{i_1})\cdots r_{i_n}(t_{i_n})}.
\end{align}
Therefore, we mainly consider correlation functions of $r_i$ instead of $r^*_i$ for the rest of this paper.

By differentiating $Z$ with respect to $b_i$ in addition to $\hat{b}$, we can formally consider the correlation function of $r_i$ and $\hat{r}_i:$
\begin{multline}
\ev{ r_{i_1}(t_{i_1})\cdots r_{i_n}(t_{i_n})
\hat{r}_{j_1}(t_{j_1})\cdots \hat{r}_{j_m}(t_{j_m})
}=\\
\fdv{\hat{b}_{i_1}(t_{i_1})} \cdots \fdv{\hat{b}_{i_n}(t_{i_n})}
\frac1i\fdv{b_{j_1}(t_{j_1})} \cdots \frac1i\fdv{b_{j_m}(t_{j_m})}
Z\eval{}_{\hat{b}_i=0}.
\end{multline}
In particular, we readily notice that the correlation functions only of $\hat{r}_i$ always vanish because
\begin{align}
\ev{\hat{r}_{j_1}(t_{j_1})\cdots \hat{r}_{j_m}(t_{j_m})}
&=\frac1i\fdv{b_{j_1}(t_{j_1})} \cdots \frac1i\fdv{b_{j_m}(t_{j_m})}
Z\eval{}_{\hat{b}_i=0}
\\&=\frac1i\fdv{b_{j_1}(t_{j_1})} \cdots \frac1i\fdv{b_{j_m}(t_{j_m})}1
\\&=0
\label{e:correl_hatr}
\end{align}
due to the normalization condition of $Z$.

Introducing the second cumulant generating function of the random variables $J_{ij}$ (see also the typical examples in Table.~\ref{t:UnivClass}),
\begin{align}
\label{e:DefKJ}
    K_J(q;\theta_a(N)) \coloneqq \log\qty[\int dJ_{ij} P_J(J_{ij};\theta_a(N))e^{iqJ_{ij}}],
\end{align}
we can rewrite the generating function Eq.~\eqref{generating fn.} as
\begin{align}
\label{generating fn2}
    Z[b,\hat{b}] = \int D\hat{r}Dr \exp(i\bar{S} + \int dt\sum_{i=1}^N \hat{b}_ir_i),
\end{align}
where the action $\bar{S}[r,\hat{r}]$ is given by
\begin{multline}
    \bar{S}[r,\hat{r}] \coloneqq
    \int_0^\infty dt \sum_{i=1}^N \hat{r}_i \qty(-\dv{t} r_i - r_i + b_i)
    \\ +
     \frac1i \sum_{i,j=1}^N K_J \qty(\int dt\, \hat{r}_i\phi(r_j)).
    \label{e:DefBarS}
\end{multline}
In the following discussion, for convenience, we denote the parameters of the probability distribution $P_J$ as $\theta_a(N)~(a=1,\ldots,\Np)$, where $\Np$ is the number of parameters of $P_J$, and express the $N$-dependence of $P_J(J_{ij})$ and $K_J$ explicitly through them.
When considering the thermodynamic limit ($N\to\infty$), we can expand $K_J$ in powers of $N$ as 
\begin{align}
    & K_J(q;\theta_a(N)) = \frac{1}{N} K(q) + \sorder{N^{-1}},
\end{align}
where $K(q)$ is defined as
\begin{align}
    & K(q) \coloneqq \lim_{N\to\infty} NK_J(q;\theta_a(N)).
\label{e:DefK}
\end{align}
We have assumed that the parameters $\theta_a(N)$ are tuned to remain the function $K(q)$ finite in the large-$N$ limit. We have shown some examples of $K(q)$ for typical distribution functions in Table.~\ref{t:UnivClass}.
The relation between the probability distribution $P_J(J_{ij})$ and the form of $K(q)$ is discussed in detail in Sec.~\ref{s:Universality}.

In the following, we neglect the sub-leading terms $\sorder{N^{-1}}$ because, as shown below, they do not contribute to the network dynamics in the $N\to \infty$ limit.

\subsection{Effective equation of motion}
Introducing the auxiliary fields $C_{n}(t_1,\cdots,t_n)$ and using the Dynamical Mean-Field approximation in the $N\to \infty$ limit, we can separate the generating function into single-node contributions, 
$Z[b,\hat{b}]\sim \prod_i Z_i[b_i,\hat{b}_i]$, and each contribution is given as follows: 
\begin{align}
\label{e:DMF_generating_fn}
&Z_i[b_i,\hat{b}_i] = \int D\hat{r}_iDr_i \exp(i\bar{S}_* + \int dt \hat{b}_ir_i),\\
&\bar{S}_* \coloneqq \bar{S}_0 +\bar{S}_C,\\
&\bar{S}_0 \coloneqq \int dt\  \hat{r}_i \qty(-\dv{t} r_i - r_i + b_i),\\
&\bar{S}_C \coloneqq \frac1i \sum_{n=1}^{\infty} \frac{\kappa_n}{n!} i^n \int [dt_j]_n C_n(t_1,\cdots,t_n)\hat{r}_i(t_1)\cdots\hat{r}_i(t_n).
\end{align}
See \app{a:DetailDerivation} for the details of the derivation.

Next, we derive the effective equation of motion of single nodes $r_i(t)$ from the generating function~Eq.~\eqref{e:DMF_generating_fn} under DMF approximation. 
Let us consider a random time series variable $\eta(t)$ which satisfies for any $n$,
\begin{equation}
\label{e:cumulant_eta}
    \ev{[\eta(t_1)\cdots\eta(t_n)]_C}_\eta=\kappa_nC^*_n(t_1,\cdots,t_n),
\end{equation}
where $[\cdots]_C$ denotes the cumulant.
$\ev{\cdots}_\eta$ denotes the average over its probability distribution $P_\eta(\eta)$, which is defined to satisfy
\begin{align}
    e^{i\bar{S}_C}
    &=\exp{\sum_{n=1}^{\infty} \frac{\kappa_n}{n!} i^n \int [dt_j]_n C^*_n\hat{r}_i(t_1)\cdots\hat{r}_i(t_n)}\\
    &=\int D\eta P_\eta(\eta) e^{i\int dt \hat{r}_i(t)\eta(t)}.
\end{align}
This definition means that we can rewrite the contribution of the action $\bar{S}_C$ as the moment generating function of $\eta$. 
Applying this formula to the Eq.~\eqref{e:DMF_generating_fn}, we obtain a renewed form of $Z_i$ with $\eta$,
\begin{multline}
\label{e:effective generating fn. with noise}
    Z_i[b,\hat{b}] 
    =
    \int D\eta D \hat{r}_iDr_i \times 
    \\
    P_\eta(\eta)\exp(i\bar{S}_0[r_i,\hat{r}_i; b_i+\eta] + \int dt\ \hat{b}_ir_i).
\end{multline}

Finally, following the reverse procedure of derivation of Eq.~\eqref{generating fn.} from Eq.~\eqref{network eq}, 
we arrive at the effective equation of motion of a single node $r_i(t)$ in the large-$N$ limit as follows:
\begin{equation}
\label{e:effective EOM}
    \dv{r_i}{t} {(t)} = -r_i(t)  + \eta(t)+ b_i(t)
\end{equation}
This result shows that, in the large-$N$ limit, the coupling term between nodes in Eq.~\eqref{network eq} is replaced by the random time series $\eta(t)$, which is generated according to the probability distribution specified by the cumulants~Eq.~\eqref{e:cumulant_eta} self-consistently. 

In particular, most previous studies have been devoted to the case where only the first and the second cumulants are non-zero, i.e.,  $\kappa_n=0$ $(n> 2)$. 
(We can find an exceptional case, for example, in Ref.~\cite{Kusmierz2020}.)
As is mentioned in \app{a:wideness}, these conditions are guaranteed in the large-$N$ limit for wide classes of probability distributions $P_J$.
In such a simple case, by deriving the self-consistent equations for the first and the second cumulants (and introducing the two-replica formalism), we can analytically estimate various dynamical quantities, such as the maximum Lyapunov exponent with and without driving signal~\cite{Sompolinsky1988, Molgedey1992,Rajan2010}, signal to noise ratio~\cite{Toyoizumi2011}, and memory curve~\cite{Schuechker2018}.
Meanwhile, in the general case, it is possible that the random network possesses an infinite number of non-zero cumulants $\kappa_n$.
As a result, we have to deal with an infinite number of self-consistent equations. 
Of course, it seems analytically intractable, and thus we have no choice but to execute some numerical analysis for this case (Sec.~\ref{s:NumCalc}).

Here we give a brief comment on the case of the discrete-time network models. 
In the context of the RC, instead of the continuous-time model~Eq.~\eqref{network eq}, the following discrete-time network model, so-called {\it Echo state network}~\cite{Jaeger2001,Jaeger2004}, is often investigated:
\begin{equation}
    r_i(T+1) = \sum_{j=1}^N J_{ij} \phi(r_j(T)) + b_i(T),
\end{equation}
where $T(\in \mathbb{Z})$ denotes the discretized time.
Even for this case, the framework discussed so far can be generalized straightforwardly, and in the end, corresponding to Eq.~\eqref{e:effective EOM}, we obtain the effective equation,
\begin{equation}
    r_i(T+1) =  \eta(T) + b_i(T),
\end{equation}
where $\eta(t)$ is the random time series variable whose cumulants are self-consistently related to the reservoir states as
\begin{equation}
    \ev{[\eta(T_1)\cdots\eta(T_n)]_C}_\eta=\kappa_nC_n(T_1,\cdots,T_n).
\end{equation}
Especially for the case where $J_{ij}$ is a Gaussian variate, the mean-field theory for these discrete-time models has already been investigated in several kinds of  literature~\cite{Molgedey1992,Toyoizumi2011,Massar2013, Haruna2019}. Qualitative differences between continuous-time models and discrete-time models have been suggested in the previous work~\cite{Schuechker2018}.
In our numerical analyses in Sec.~\ref{s:NumCalc}, however, we focus primarily on the continuous-time model, since we have numerically confirmed that the differences between continuous-time models and discrete-time models seems not to be essential for our numerical results from a qualitative perspective.

Before moving on to the next section, we explain why $K(q)$ should take a non-zero finite value as $N\to\infty$, from the point of view of the effective equation of motion.
Let us consider the case where $K(q)$ becomes zero ($\sorder{N^0}$).
In this case, $\kappa_n$ becomes $\sorder{N^0}$ and vanish in the limit,
and then, all time-correlation functions of $\eta(t)$ also become zero from \siki{e:cumulant_eta}.
These conditions can be achieved only when $\eta(t)$ is identically zero for arbitrary time.
Because the effective equation of motion depends on the probability distribution $P_J$ only through $\eta(t)$, the dynamics are not changed by choice of $P_j$.
Therefore, we should tune parameters for $K$ to be non-zero in the limit so that the probability distribution has non-negligible effects on the averaged dynamics.

That is, the dynamics of the RCs is mostly controlled by the external input $b_i(t)$ and sub-leading terms mentioned above appears as small perturbations to $r_i(t)$ with small amplitudes in the order $\order{N^{-1}}$.
Then, each node behaves as single independent damping system under the external input $b_i(t)$ in the case of $K(q)=0$.
This leads to that the parameters of the probability function of the coupling $J_{ij}$ have only tiny effects on the RC performance.

How about the case where $K(q)$ becomes divergent?
In this case, each $\kappa_b$ in \siki{e:DefSbar} is not finite ($\order{N^0}$) but divergent, and all time-correlation functions of $\eta$ becomes divergent.
In the case where the amplitude of the external input $b_i(t)$ does not depend on the network size $N$ and has finite ($\order{N^0}$) values in the limit, $\eta(t)$ takes very large values compared to it.
Therefore, the terms $b_i(t)$ in the effective equation of motion (\siki{e:effective EOM}) can be neglected.
Actually, the case of $K(q)\sim\mathcal{O}(N)$ corresponds to a very limited region of the parameter space in the Fig.~6-7, such as $1/J\to 0$ or $J_0/J\to \infty$, where the computational performance is typically not good.
This argument shows that we should keep $K(q)$ finite to achieve the RC scheme.

The above arguments show the necessity of tuning to retain $K(q)$ non-zero and finite ($\order{N^0}$) as $N\to \infty$.
In the next section, we investigate the resultant forms of $K(q)$ by this parameter-tuning and find that they are classified into several ``universality" classes.

\section{Universality}
\label{s:Universality}
In this section, we discuss the large-$N$ behavior of the effective action, especially the cumulant generating function term $K(q)$.
As was shown in the previous sections, only $K(q)$ controls the dynamics of random networks in the $N\to\infty$ limit, and thus we should tune the parameter of the probability distribution $P_J(J_{ij})$ to remain $K(q)$ finite.
Intriguingly, it is clarified that various probability distributions fall into the same form of $K(q)$, and thus, a set of the corresponding networks shows the common dynamical properties in the thermodynamic limit.
This result means that the dynamics of the random network models can be classified into several universality classes, according to the large-$N$ behaviors of the generating functions of the probability distributions of the coupling constants.

\subsection{Large-N behavior of K}
\label{s:LargeNbehaviourK}
Before entering the general discussion, we give a simple example where $P_J(J_{ij})$ equals the Gaussian distribution with variance $\sigma^2$ and zero mean.
Many previous studies~\cite{Sompolinsky1988, Toyoizumi2011, Rajan2010, Crisanti2018, Schuechker2018} have considered this case.
The cumulant generating function $K_J(q)$ is given by
\begin{align}
K_J(q) = -\sigma^2q^2/2
\end{align}
and hence $K(q)$ is calculated as
\begin{align}
    K(q) = \lim_{N\to\infty} NK_J(q) = -\lim_{N\to\infty} N\sigma^2q^2/2.
\end{align}
In order to remain $K(q)$ finite as $N\to\infty$, we should set the $N$-dependence of the parameter $\sigma(N)$ as $\sigma(N)^2=\bar{\sigma}^2/N$, where $\bar{\sigma}^2$ is a $N$-independent parameter.
Finally we obtain the finite $N$-independent function $K(q)$ as
\begin{align}
K(q) = -\bar{\sigma}^2q^2/2
\end{align}
for a Gaussian distribution with zero mean. 
We generalize the above procedure for an arbitrary distribution and discuss how to decide the $N$-dependence of the parameters $\ta_a(N)$.

First of all, we should identify the values which $\ta_a(N)$ approach as $N\to\infty$.
These values can be readily found from the definition of $K(q)$ (\siki{e:DefK}).
When $K(q)$ remains finite in the large-$N$ limit ($K(q)=\order{N^0}$), it follows from the definition of $K(q)$ that the cumulant generating function $K_J(q)$ should be represented, up to the $N$-leading order, as 
\begin{align}
\label{e:K-limit}
    K_J(q;\ta_a(N)) = \frac{1}{N} K(q) + \sorder{N^{-1}},
\end{align}
which leads to 
\begin{align}
    \lim_{N\to \infty} K_J(q;\ta_a(N)) = K_J(q;\ta_a(\infty)) = 0.
    \label{e:ZeroPoint}
\end{align}
This equation means that we should tune parameters to approach zeros of $K_J$ as $N\to\infty$.

Generally, zero set of the second cumulant generating function $K_J(q;\ta_a)$ can be decomposed into several manifolds in the parameter space spanned by $\{\ta_a(N)\}_{a=1,\cdots,N_{para}}$.
(Mathematically, the zero set is represented as a semi-algebraic set in this setting.
The semi-algebraic set consists of its irreducible components, and each component is referred to as a ``manifold" here.) 
The way of the parameter-tuning depends on a choice of the manifold which $K_J(q)$ approaches as $N\to\infty$, and so does the form of $K(q)$ as a result.
Each of these manifolds has a fixed dimension $\Nf$ defined as the number of parameters that can have arbitrary values on the manifold.
These parameters are not needed to tune as $N\to\infty$.
However, the other parameters should be controlled to approach asymptotically to the manifold if we want to remain $K(q)$ finite.
The number of these parameters $\Nt$ equals the codimension of the manifold, that is, $\Nt=\Np-\Nf$. 
Note that the value of $\Nf$ and $\Nt$ differs between the manifolds, even for the same distribution.

As a demonstration, we give a simple example of the above argument.
Let us consider the Gamma distribution, whose second cumulant generating function is given by $K_J(q;\theta,k) = -k \log(1-i\theta q)$, where $k$ and $\theta$ are non-negative real parameters.
The set of paramteres $\{\ta_a\}$ is given by $\{\ta,k\}$ in this case.
The zeros of $K_J(q;\ta,k)$ in the $\theta$-$k$ space are (1) $\{(\ta, 0) \,|\, \ta \in \mathbb{R}_{+} \}$ and (2) $\{(0, k) \,|\, k \in \mathbb R_{+} \}$.
Therefore, we should tune just $k$ in the case (1) and $\theta$ in the case (2) in the large-$N$ limit.
The detailed way of the parameter-tuning in each case and the corresponding universality class is given later in \Sec{s:gamma}.

Let us determine the $N$-dependence of the parameters $\{\ta_a(N)\}$.
We give a {\it practical} way of the tuning, which is sufficient in most cases.
(In \app{a:renormalization}, we discuss it more precisely with the renormalization group method, which is often used in studies of quantum field theories in condensed matter or high-energy physics.)
In the following, we choose one of the zeros of $K_J(q)$, and denote parameters which are needed to tune as $\vec{\za} = (\zeta_a)~(a=1,\ldots,\Nt)$, and ones which are not needed to as $\vec{\xi}=(\xi_a)~(a=1,\ldots,\Np-\Nt)$.
Remember that $\{\ta_a\} = (\vec\zeta, \vec\xi)$ holds.

For simplicity, suppose that $P_J$ can be described with $\Nt$-th order Taylor polynomial, in other words, its cumulants are well-defined up to at least $\Nt$-th order.
Then, $K_J$ is $\Nt$-th-order differentiable at $q=0$ and can be expanded as
\begin{align}
    K_J(q;\vec{\za},\vec{\xi}) = \sum_{b=1}^{\Nt} \kappa_b(\vec{\zeta},\vec{\xi})\frac{(iq)^b}{b!} 
    + R(q;\vec\za,\vec\xi),
\end{align}
where $R(q)$ is the so-called Peano form of the remainder and has the order of $\sorder{q^{\Nt}}$ near $q=0$.
$\kappa_b$ is the $b$-th cumulant, which is a function of $\vec\za$ and $\vec\xi$.
Therefore, $K(q)$ is given by
\begin{multline}
\label{e:KLimit}
    K(q) = 
    \lim_{N\to\infty} \sum_{b=1}^{\Nt}
    \qty( N\kappa_b(\vec{\zeta}(N),\vec{\xi})\frac{(iq)^b}{b!})
    \\
    + NR(q;\vec{\zeta}(N),\vec{\xi}).
\end{multline}
From this expression, it is obvious that $\vec\za(N)$ should satisfy that 
\begin{align}
    \kappa_b(\vec{\zeta}(N),\vec{\xi}) = \frac{\bar{\kappa}_b}{N}.
    \label{e:ScaleMu}
\end{align}
$\bar{\kappa}_b$ is introduced as a $N$-independent parameter and controls the form of $K(q)$.

To determine $\vec{\zeta}(N)$ concretely, let us consider the map 
$\vec{F}_\kappa:\mathbb{R}^{\Nt} \ni \vec{\za} \mapsto \vec{\kappa} \in \mathbb{R}^{\Nt} $ and its inverse map $\vec{F}^{-1}_\kappa$, where $\vec{\kappa}=\{\kappa_{a=1,\ldots,\Nt}\}$.
More specifically, $\vec{F}_\kappa$ is given by
\begin{align}
    \vec{F}_\kappa (\vec{\za}) \coloneqq \vec{\kappa}(\vec{\za},\vec{\xi})
\end{align}
for each fixed $\vec{\xi}$.
Then, to satisfy \siki{e:ScaleMu}, $\vec{\zeta}(N)$ is determined as
\begin{align}
    \vec{\zeta}(N) = \vec{F}_{\kappa}^{-1}\qty(\frac{\vec{\bar{\kappa}}}{N}).
    \label{e:NDepPara}
\end{align}
Here, $\vec{\bar{\kappa}}=\{\bar{\kappa}_{a=1,\ldots,\Nt}\}$.
We emphasize that $\vec\xi$ does not need to be tuned and is a parameter of $K(q)$ as well as $\vec{\bar{\kappa}}$.

One may worry about finiteness of the term $\lim_{N\to\infty} NR(q;\theta_a(N))$ in \siki{e:KLimit}.
Of course this term may not be finite in some distributions;
however this point does not matter practically.
This is because second cumulant generating function $K_J$ is first-order differentiable at $\vec{\kappa}=0$ in most cases.
When $K_J$ is first-order differentiable, $K(q)$ always becomes finite and is linear in $\vec{\bar{\kappa}}$.
Under this assumption, $K_J$ is expanded as 
\begin{align}
    K_J = \sum_{b=1}^{\Nt} \eval{\pdv{K_J}{\kappa_b}}_{\vec{\kappa}=0}\kappa_b + \sorder{\kappa_b}
\end{align}
as $\kappa_b \ll 1$.
Therefore, with \siki{e:ScaleMu}, we get $K(q)$ as
\begin{align}
    K &= \lim_{N\to\infty} 
    \qty(
    \sum_{b=1}^{\Nt} \bar{\kappa}_b \eval{\pdv{K_J}{\kappa_b}}_{\vec{\kappa}=0} + \sorder{N^{0}}
    )
    \\ & = \sum_{a=1}^{\Nt} \bar{\kappa}_b \eval{\pdv{K_J}{\kappa_b}}_{\vec{\kappa}=0}.
    \label{e:DifferentiableK}
\end{align}
It should be noted that from \siki{e:DifferentiableK}, we find that $K$ is linear function of $\vec{\bar{\kappa}}$ and the corresponding probability distribution has reproducibility about $\vec{\bar{\kappa}}$:
\begin{align}
    K(q;\vec{\bar{\kappa}}) + K(q;\vec{\bar{\kappa}}') = K(q;\vec{\bar{\kappa}}+\vec{\bar{\kappa}}').
\end{align}

This tuning method can be also generalized to distributions which do not have finite cumulants, such as the Cauchy distribution. 
The second cumulant generating function of a general probability distribution can be expanded with some linearly-independent functions $f_k(q,\vec{\xi})$ as
\begin{align}
    K(q) = \sum_{k=1}^\infty \Lambda_k(\vec{\zeta}) f_k(q,\vec{\xi}).
\end{align}
where each $\Lambda_k$ is a expansion coefficient.
The linearly-independence means that $K(q;\vec{\za}(\infty),\xi_a)=0$ for an arbitrary value of $q$ is equivalent to $\Lambda_k(\vec{\za}(\infty))=0$.
Then in analogy with the case of distributions with finite cumulants (\siki{e:ScaleMu}), it is sufficient to tune parameters so that
\begin{align}
    \Lambda_b(\vec{\zeta}(N)) = \frac{\bar{\Lambda}_b}{N}.
\end{align}

Let us explain the above procedure with the Cauchy distribution, whose second cumulant generating function is given by
\begin{align}
     K_J(q;\delta,\gamma) = i\delta q - \gamma \abs{q}.
\end{align}
It is natural to choose $f_1(q)=iq$ and $f_2(q)=-\abs{q}$ as the linearly-independent functions.
Because $\Lambda_1=\delta$ and $\Lambda_2=\gamma$, the zero of $K_J$ is given by $\delta=\gamma=0$.
Note that because $\vec{\Lambda}(\vec{\zeta})=\vec{\zeta}=(\delta,\gamma)$ holds, their inverse function $\vec{F}^{-1}_\kappa(\vec{\Lambda})$ is given by $\vec{F}^{-1}_{\kappa}(\vec{\Lambda})=\vec{\Lambda}$.
Then, the $N$-dependence of $\delta$ and $\gamma$ is determined as
\begin{align}
    \delta &
    = \qty(\vec{F}^{-1}_\kappa\qty(\frac{\vec{\bar{\Lambda}}}{N}))_1 
    = \frac{\bar{\Lambda}_1}{N},\\
    \quad 
    \gamma &
    = \qty(\vec{F}^{-1}_\kappa\qty(\frac{\vec{\bar{\Lambda}}}{N}))_2
    = \frac{\bar{\Lambda}_2}{N},
\end{align}
where each $\bar{\Lambda}_a$ is a $N$-independent parameters.

Although the choice of linearly-independent functions $f_k$ has ambiguity, our intention here is to give just a practical way of determining how to tune parameters, rather than a perfect one.
As far as we have investigated, it is sufficient to choose them so that the numbers of $f_k$ are minimized.

Before explaining each universality class, it is instructive to comment on the difference between our tuning here and the (generalized) central limit theorem.
The limit we have considered so far is 
\begin{align}
\label{e:OurLim}
    \lim_{N \to \infty} N K_J(q;\vec{\ta}(N))
\end{align}
with {\it $N$-dependent} parameters $\ta_a(N)$, whereas the one discussed in the (generalized) central limit theorem is 
\begin{align}
\label{e:CentralLim}
    \lim_{N \to \infty} N K_J\qty(\frac{q}{\sqrt{N}};\vec{\ta}) 
\end{align}
with {\it $N$-independent} parameters $\ta_a$.
We emphasize that these limits are {\it different}.
In fact, we get other forms of $K(q)$ than the Gaussian or stable distributions, which are conclusions of the (generalized) central limit theorem.
Because these limits have been sometimes confused in some previous literatures, we should be careful about this difference.

\subsection{Universality Class}
\label{s:ExamUC}
In this section, we give some examples of universality classes.
Table \ref{t:UnivClass} shows the summary of these examples.
It should be emphasized that the following classes are just examples and that other classes are possible to be realized if we consider more complex distribution functions than those discussed below.

\subsubsection{Delta Class}

Let us start to discuss the following form of $K(q)$:
\begin{align}
    K(q) = i\bar{\mu} q.
\label{e:KDelta}
\end{align}
We notice readily that This is the same as $K_J(q)$ of the Dirac delta distribution with the mean $\bar{\mu}$.
For this reason, we refer to distributions that have the above form of $K(q)$ as the ``delta" class.

For example, the exponential and Pareto distribution are included in this class when parameters are tuned to satisfy
\begin{align}
    E[J_{ij}] = \frac{\bar{\mu}}{N}.
\end{align}
If we tune $\theta$ in the Gamma distribution as $\theta = \bar{\theta}/N$ and let $k$ be an $N$-independent constant, then we also get \siki{e:KDelta} with $\bar{\mu} = k \bar{\theta}$.
This is the case (2) of the Gamma distribution in \Sec{s:LargeNbehaviourK}.

As another example, the Log-normal distribution belongs to this class.
Its second cumulant generating function is given by 
\begin{align}
    K_J(q) = \log( 1 + \sum_{n\geq 1} \frac{(iq)^n}{n!} \exp(n\mu+n^2\frac{\sigma^2}{2}) ).
    \label{e:KLognormal}
\end{align}
The zero of $K_J$ is $\exp(\mu)=0$, that is, $\mu = -\infty$.
Note that $\exp(\sigma^2)\geq 1$ for any real $\sigma$ and cannot become zero.
The leading-order term of $K_J(q)$ around $\exp(\mu) = 0$ is
\begin{align}
    K_J(q) = iq \exp(\mu+\sigma^2) + \order{\exp(2\mu)}.
\end{align}
Therefore we should tune only $\mu$ so that $N\exp(\mu)$ is finite, as 
\begin{align}
    N \exp(\mu) = \exp(\mu'),
\end{align}
that is,
\begin{align}
    \mu = -\log N + \mu',
\end{align}
where $\mu'$ is independent of $N$.

Then, $K$ is given by 
\begin{align}
    K(q) = iq \exp(\mu'+\sigma^2),
\end{align}
which corresponds to $K$ of Dirac delta distribution with the mean  $\bar{\mu}=\exp(\mu'+\sigma^2)$.

\subsubsection{Gauss Class}
We introduce the ``Gauss class" here, whose $K(q)$ is given by
\begin{align}
    K(q) = i \bar{\mu} q - \frac{\bar{\sigma}^2}{2} q^2.
    \label{e:KGauss}
\end{align}
This corresponds to the second cumulant generating function of the Gaussian distribution with the mean $\bar{\mu}$ and variance $\bar{\sigma}^2$.
Although \siki{e:KGauss} has a very simple form, quite many distribution that have two or more parameters belongs to this class when we tune their parameters to satisfy
\begin{align}
\label{e:GaussAssum}
    E(J_{ij}) = \frac{\bar{\mu}}{N},\ \  V(J_{ij}) = \frac{\bar{\sigma}^2}{N}.
\end{align}
We confirm that the uniform, triangular, Gumbel and Laplace distribution also belong to this class.
In fact, most of previous studies has been devoted to the analysis of this class with $\bar{\mu}=0$~\cite{Sompolinsky1988,Rajan2010,Schuechker2018,Crisanti2018}.

Actually this class is very wide, and in \app{a:wideness}, we show that a distribution belongs to this class if it satisfies some reasonable conditions.

\subsubsection{Stable Class}
The ``Stable class" is given by
\begin{align}
    K(q) = i\bar{\delta}q - \bar{\gamma} \abs{q}^\alpha \qty(1+i\beta \sgn(q) \omega(q,\alpha)),
\end{align}
where $\bar{\delta}, \bar{\gamma}, \alpha$ and $\beta$ are arbitrary parameters.
$\omega(q,\alpha)$ is defined as
\begin{empheq}[left={\omega(q,\alpha)\coloneqq\empheqlbrace}]{align}
 \tan \qty(\frac{\pi \alpha}{2})\quad & (\alpha \neq 1), \\
 \frac{2}{\pi} \log \abs{q}\quad & (\alpha = 1) .
\end{empheq}
This $K(q)$ corresponds to $K_J(q)$ of the stable distribution with the parameters of 
\begin{align}
    \delta = \bar{\delta},\ \  \gamma = \bar{\gamma}.
\end{align}

It should be noted that the Gauss class is a special case of the stable class with identification of $\bar{\delta}=\bar{\mu},\bar{\gamma}=\bar{\sigma}^2/2$ and $\alpha=2$.

\subsubsection{Gamma Class}
\label{s:gamma}
Next, we explain the ``Gamma class".
$K(q)$ of this class is given by
\begin{align}
    K(q) = -\bar{k}\log(1-i\theta q),
\end{align}
where $\bar{k}$ and $\theta$ are arbitrary parameters.
This $K(q)$ corresponds to the limit of $K_J(q)$ for the Gamma distribution with 
\begin{align}
    k = \frac{\bar{k}}{N}
\end{align}
and an arbitrary constant $\theta$, which was discussed as case (1) in \Sec{s:LargeNbehaviourK}.

This class is peculiar to the limit \siki{e:OurLim} considered in this study.
If we consider the limit (\ref{e:CentralLim}), we obtain $K(q)$ of the Gaussian distribution, not the one given above.

\subsubsection{Symmetrized Gamma Class}
\label{s:sgamma}
The last example is ``Symmetrized Gamma class",
whose $K(q)$ has the following form:
\begin{align}
\label{e:Ksgamma}
    K(q) = i\bar{\mu}q-\frac{\bar{k}}{2}\log(1-i\theta q)(1+i\theta q^*),
\end{align}
where $\bar{\mu}$, $\bar{k}$ and $\theta$ are arbitrary parameters.
We get this class by symmetrizing the probability distribution function of the Gamma distribution and shifting the distribution by $\mu$.
That is, the probability distribution function 
\begin{align}
    P_J(x) = \frac{\abs{x-\mu}^{k-1} \exp(-\abs{x-\mu}/\theta)}{\Gamma(k)\theta^k}\quad  (x \in \mathbb{R})
\end{align}
realizes the above $K(q)$ by tuning $\mu$ and $k$ as
\begin{align}
    \mu = \frac{\bar{\mu}}{N}, \quad k = \frac{\bar{k}}{N},
\end{align}
with $\theta$ fixed.

\begin{turnpage}

\renewcommand{\arraystretch}{2.5}
\begin{table*}[htb]
  \begin{tabular}{|c|c|c|c|c|}
  \hline
  Distribution & $P_J(x)$ & Parameter scaling & $\log K_J(q)$ & $K(q)$ 
  \\\hline\hline
  Delta & $\delta(x-\mu)$ & $\displaystyle \mu=\frac{\bar{\mu}}{N}$ & $i\mu q$ & $i\bar\mu q$
  \\\hline
  Exponential & $\lambda \exp(-\lambda x)$ & $\displaystyle \lambda = \frac{N}{\bar{\mu}}$ & $-\log(1-iq\lambda^{-1})$ & $i\bar\mu q$
  \\\hline
  Log-normal & $\displaystyle \frac{1}{\sqrt{2\pi}\sigma x}\exp(-\frac{(\log x-\mu)^2)}{2\sigma^2})$ & $\displaystyle \mu=\log \frac{\bar{\mu}}{N}- \sigma^2$ &     
  $ \displaystyle \log\qty(\sum_{n=0}^{\infty} \frac{(it)^n}{n!} e^{ n\mu+n^2\frac{\sigma^2}{2}})$ & $i\bar\mu q$
  \\\hline\hline
  Normal & $\displaystyle \frac{1}{\sqrt{2\pi\sigma^2}}\exp(-\frac{(x-\mu)^2}{2\sigma^2})$ & $\displaystyle \mu=\frac{\bar\mu}{N},\  \sigma= \frac{\bar\sigma}{\sqrt N}$ &$\displaystyle i\mu q - \frac{\sigma^2q^2}2$ & $\displaystyle i\bar\mu q - \frac{\bar\sigma^2q^2}{2}$ 
  \\\hline
  Uniform & $ \displaystyle \frac{\chi_{[a,b]}}{b-a}$ &  $ \displaystyle b=\frac{\bar{\mu}}{N} + \sqrt{\frac{3}{N}}\bar\sigma$, $\displaystyle a=\frac{\bar{\mu}}{N} - \sqrt{\frac{3}{N}}\bar\sigma$ & $iq\mu + \log(\mathrm{sinc}(\sqrt{3}\sigma q))$ & $\displaystyle i\bar\mu q - \frac{\bar\sigma^2q^2}{2}$ 
  \\\hline
  Gumbel & $\displaystyle \exp(-\exp(-\frac{x-\mu}{\eta}))$ & $\displaystyle \mu = \frac{\bar{\mu}}{N} - \gamma\sqrt{\frac{6}{\pi^2 N}} \bar{\sigma}$, $\displaystyle \eta = \sqrt{\frac{6}{\pi^2N}} \bar{\sigma}$ & $i\mu q + \log\Gamma(1-i\eta q)$ & $\displaystyle i\bar\mu q - \frac{\bar\sigma^2q^2}{2}$ 
  \\\hline\hline
  Stable & (undefined) & $\delta=\bar{\delta}$, $\gamma=\bar{\gamma}$ & $i{\delta}q - {\gamma}\abs{z}^\alpha\sgn(z) \omega(z,\alpha)$ & $i\bar{\delta}q - \bar{\gamma}\abs{z}^\alpha\sgn(z) \omega(z,\alpha)$
  \\\hline\hline
  Cauchy & $\displaystyle \frac{1}{\pi} \frac{\gamma}{(x-x_0)^2+\gamma^2}$ & $\displaystyle x_0= \frac{\bar{x}_0}{N}$, $\displaystyle  \gamma=\frac{\bar{\gamma}}{N}$ & $ix_0q -\gamma|q|$ & $i\bar{x}_0q - \bar\gamma|q|$
  \\\hline\hline
  Gamma & $\displaystyle \frac{x^{k-1} \exp(-x/\theta)}{\Gamma(k)\theta^k}$ & $\displaystyle k=\frac{\bar{k}}{N}$ & $-k\log(1-i\theta q)$ & $-\bar{k} \log\qty(1-i\bar{\theta} q)$
    \\\hline\hline
    \renewcommand{\arraystretch}{1}
    \begin{tabular}{c}
  symmetrized-
  \\Gamma 
    \end{tabular}
    \renewcommand{\arraystretch}{2.1}
  & $\displaystyle \frac{|x-\mu|^{k-1} \exp(-\abs{x-\mu}/\theta)}{\Gamma(k)\theta^k}$ & $\displaystyle \mu=\frac{\bar{\mu}}{N}$, $\displaystyle k=\frac{\bar{k}}{N}$ & 
   $\exp(i\mu q)\times \Re\qty[\exp(-k\log(1-i\theta q))]$ 
  & $\displaystyle i\bar{\mu}q-\frac{\bar{k}}{2}\log(1-i\theta q)(1+i\theta q^*)$
  \\\hline
  \end{tabular}
  \caption{Typical examples of relations between distribution functions $P_J(J_{ij})$ and the corresponding form of $K(q)$. 
In the column ``Parameter scaling", we show how to scale the parameters in the large-$N$ limit. $\chi_{[a,b]}$ appeared in the probability distribution function of the uniform distribution is the indicator function.}
\label{t:UnivClass}
\end{table*}
\end{turnpage}

\clearpage

\subsection{Correspondence between universality class and eigenvalue spectrum}

\begin{figure*}[t]
  \centering
  \begin{tabular}{lll}
(a)  & (b) & (c) \\
     \includegraphics[keepaspectratio, height=0.33\linewidth]
      {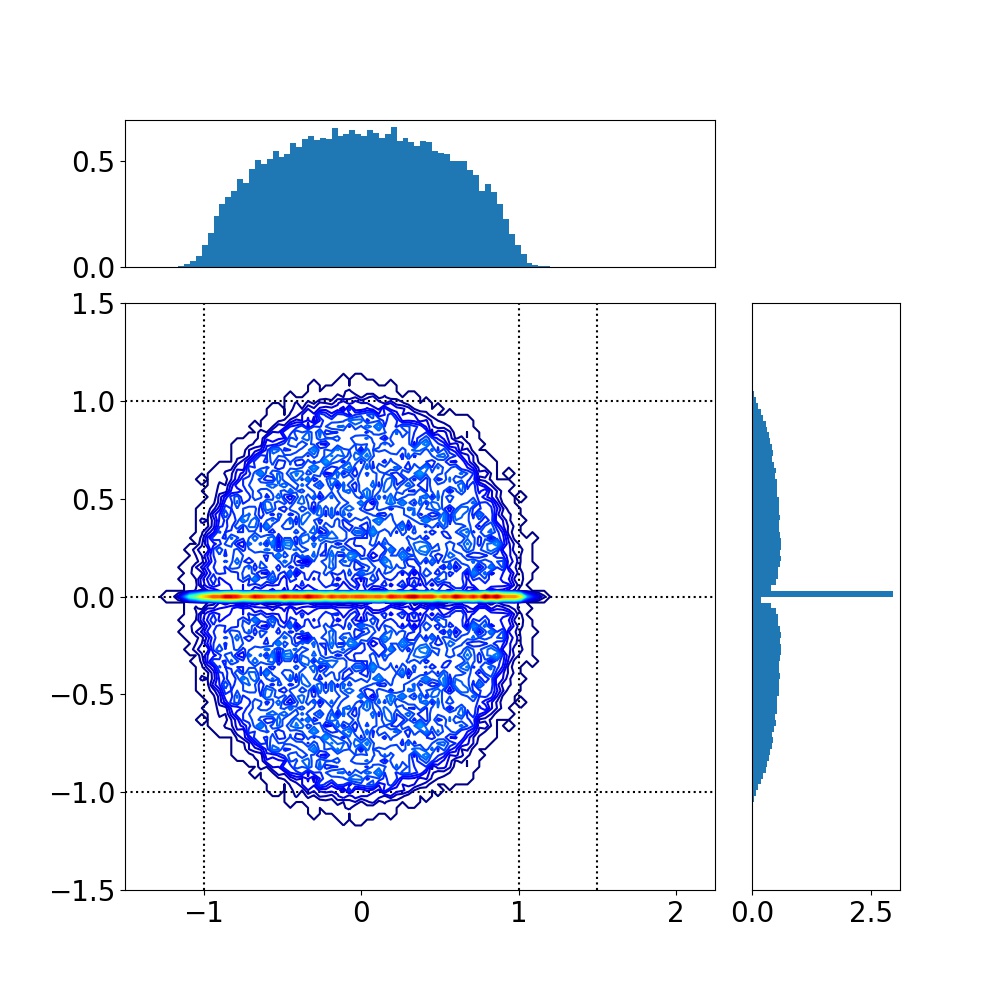} 
      &
    \includegraphics[keepaspectratio, height=0.33\linewidth]
      {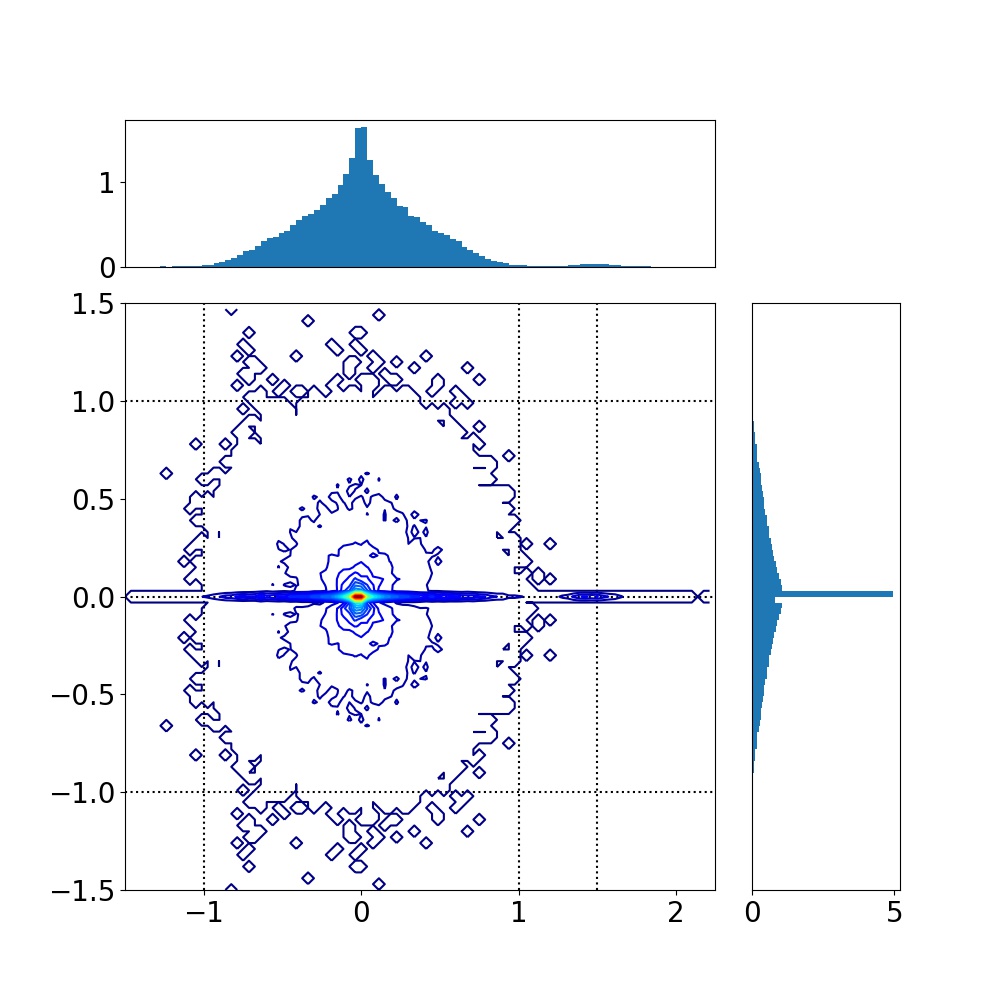} 
      &
    \includegraphics[keepaspectratio, height=0.33\linewidth]
      {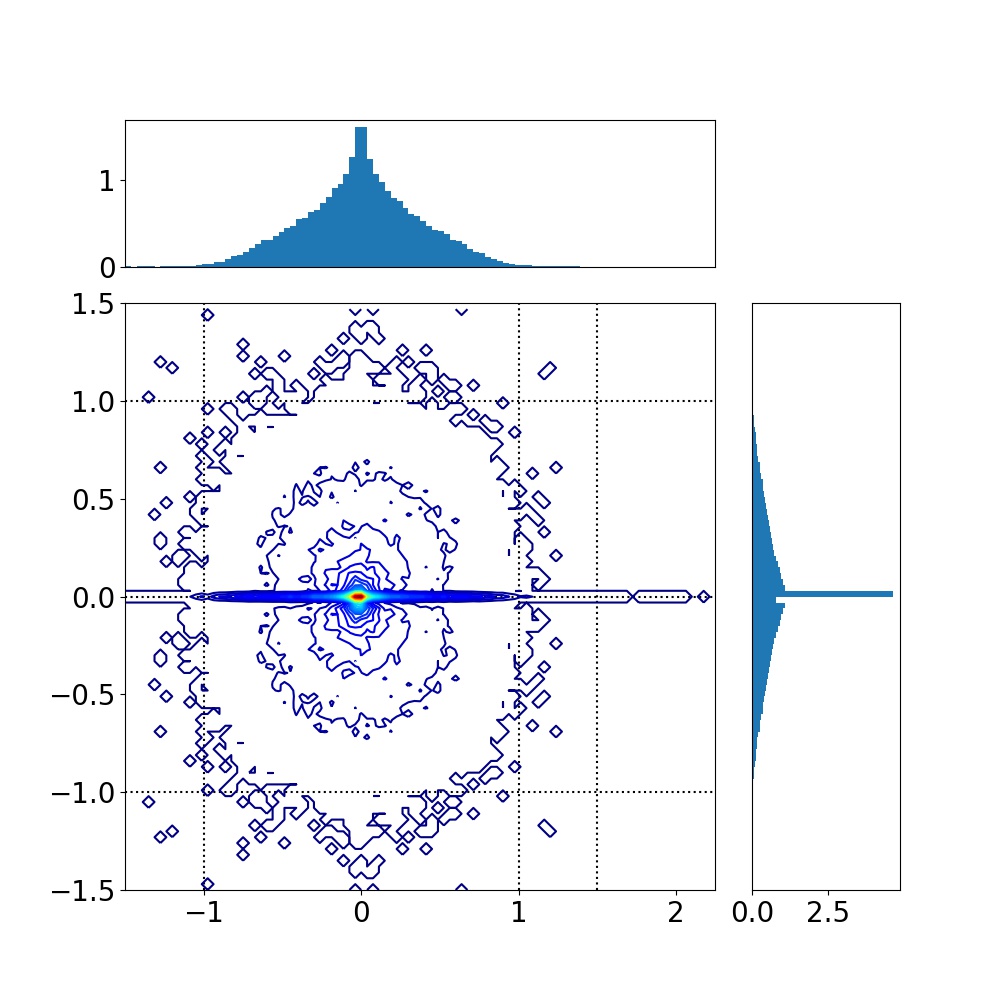} 
  \end{tabular}
 \caption{Eigenvalue spectrum density of the random coupling matrix on the complex plane. Each random matrix is sampled from (a) the Gaussian distribution (b) the Gamma distribution (c) the symmetrized-Gamma distribution with $J_0/J=1.5,1/J=1,N=500$, and the mean of the symmetrized-Gamma distribution is set to be zero.
 Each of the figures in the center of (a-c) is a color map of the eigenvalue spectrum density $\rho(\omega)$.
 The upper and right figures are integrated density in the directions of imaginary and real axis, respectively.
 }
\label{f:EigenvalueSpectrum}
\end{figure*}

In this section, we show that, in the large-$N$ limit, the eigenspectrum of random matrix $J$ is uniquely determined by specifying the universality class to which the probability distribution $P_J$ belongs.

According to the analogy between the eigenspectrum of random matrix and electrostatics discussed in Ref.~\cite{Sommer1988}, the number density $\rho(\omega)$ ($\omega=x+iy$) of eigenvalues of random matrix $J$ is given by
\begin{align}
    \rho(w) = -\frac{1}{4\pi} \qty(\pdv[2]{x}+\pdv[2]{y})\Psi(w).
\end{align}
Here $\Psi$ is the electrostatic potential in two-dimensional space $(x,y) \in \mathbb{R}^2$ and defined as
\begin{multline}
    \Psi(w) = 
    \frac{1}{N}
    \int [dJ_{ij}] 
    \times 
    \\
    \prod_{i,j}P(J_{ij})
    \log \det[(w^*E_N-J^T)(wE_N-J)],
    \label{e:DefPsi}
\end{multline}
where $E_N$ is the $N\times N$ unit matrix.

In the following, we show that this potential function $\Psi$ depends on $P_J$ only through $K(q)$ in the large-$N$ limit.
With the Gaussian integral formula, we can rewrite \siki{e:DefPsi} as
\begin{multline}
N\Psi = \int [dJ_{ij}] \prod_{i,j}P(J_{ij}) \times 
    \\
    \log\qty( \int [d^2z_i] \exp[
    -z_i^*(w^*\delta_{ij}-J_{ji})(w\delta_{jk}-J_{ij})z_k]),
\end{multline}
where $[d^2z_i] \coloneqq \prod_i d^2z_i/\pi$.

To evaluate the integration over $J_{ij}$, let us use the replica method to remove the logarithm.
We define $\Psi^{(R)}$ as
\begin{multline}
N\Psi^{(R)} = \int [dJ_{ij}] \prod_{i,j}P(J_{ij}) \times 
    \\
    \qty( \int [d^2z_i] \exp[
    -z_i^*(w^*\delta_{ij}-J_{ji})(w\delta_{jk}-J_{jk})z_k])^R.
\end{multline}
After calculating $\Psi^{(R)}$, we analytically continue 
% $R$ in 
$\Psi^{(R)}$ from $R \in \mathbb{N}$ to $R \in \mathbb{R}$, 
then the original $\Psi$ is calculated from
\begin{align}
    \Psi = \lim_{R\to 0} \frac{\Psi^{(R)}-1}{R}.
\end{align}

It is sufficient to our purpose to show that $\Psi^{(R)}$, instead of $\Psi$, depends on $P_J$ only through $K(q)$ in the large-$N$ limit. 
$\Psi^{(R)}$ is calculated as
\begin{align}
% \\
&
\begin{multlined}
N\Psi^{(R)} = \int [dJ_{ij}] [d^2z_i^{(r)}] \prod_{i,j}P(J_{ij}) \times 
    \\
    \exp[
    -z^{(r)*}_i(w^*\delta_{ij}-J^{ji})(w\delta_{jk}-J_{jk})z_k^{(r)}
    ]
\end{multlined}
\\&
\begin{multlined}
= \int [dJ_{ij}] [d^2z_i^{(r)}] [d^2y_i^{(r)}] \prod_{i,j}P(J_{ij}) 
\times  \\
    \exp
    \biggl[ -y_i^{(r)*}y_i^{(r)} 
    - i y_j^{(r)*}(w\delta_{jk}-J_{jk})z_k^{(r)} 
    \\
    - i z^{(r)*}_i(w^*\delta_{ij}-J_{ji})y_j
    \biggr]
\end{multlined}
\\&
\begin{multlined}
= \int [d^2z_i^{(r)}] [d^2y_i^{(r)}] \times  \\
    \exp
    \biggl[ -y_i^{(r)*}y_i^{(r)} 
    - i  w y_i^{(r)*} z_i^{(r)} - i w^* z^{(r)*}_iy_i^{(r)}
    \\
    + \sum_{i,j} K_J\qty(y_i^{(r)*} z_j^{(r)} + y_i^{(r)} z^{(r)*}_j)
    \biggr].
\end{multlined}
\end{align}
We have used the Gaussian integral formula to introduce the integration variables of $y_i^{(r)}$s in the second equality and executed the integration over $J_{ij}$ in the  third equality.

Let us introduce auxiliary fields in the same way as we did in \Sec{s:AuxFields}.
$K_J$ can be expanded as $K_J = K/N + \order{N^{-2}}$ in the large-$N$ limit, and then $K/N$ is expanded as
\begin{align}
    &\frac{1}{N}\sum_{j} K\qty(y_i^{(r)*} z_j^{(r)} + y_i^{(r)} z^{(r)*}_j)
    \\
    &= \frac{1}{N}\sum_{n}\frac{\kappa_{n}}{n!} i^{n} \sum_j \qty(y_i^{(r)*} z_j^{(r)} + y_i^{(r)} z^{(r)*}_j)^n
    \\
    &
    \begin{multlined}
    = \frac{1}{N}\sum_{n}\frac{\kappa_{n}}{n!} i^{n} \sum_{a=0}^n\sum_{j=1}^N 
    \binom{n}{a} (y_i^{(r)*})^a (z_j^{(r)})^a  
    \times
    \\
    (y_i^{(r)})^{n-a} (z^{(r)*}_j)^{n-a}.
    \end{multlined}
\end{align}
We also use the following identity:
\begin{align}
    1 &= \int dC_n^a \delta \qty(
    C_n^a - \frac{1}{N} (z_i^{(r)})^a (z_i^{(r)*})^{n-a}
    )
    \\
    &
    \begin{multlined}
    = \frac{N}{2\pi}\int d\widehat{C}^a_n dC^a_n \\
    \exp\qty[-i \widehat{C}^a_n
          \qty(
            C_n^a - \frac{1}{N} (z_i^{(r)})^a (z_i^{(r)*})^{n-a}
            )
        ].
    \end{multlined}
\end{align}
Performing these transformations, finally we obtain the averaged generating function and the effective action in the following form:
\begin{align}
    \Psi^{(R)} = \int \prod_{n=1}^{\infty}\prod_{a=1}^{n} DC^a_nD\widehat{C}^a_n e^{-N I_1}.
\end{align}
$I_1$ is defined as
\begin{multline}
\label{e:DefI1}
    I_1[C,\widehat{C}] 
    \coloneqq 
    -
    \frac{1}{N} \sum_{i=1}^N \log
    \qty[
        \int [dz_idy_i]
        e^{i\bar{I}_1[y_i,z_i, C, \widehat{C}]
        }
        ]
    \\
    +i \sum_{n=1}^{\infty}\sum_{a=1}^n  C_n^a\widehat{C}_n^a,
\end{multline}
where $\bar{I}_1$ is given by
\begin{multline}
\label{e:DefI1bar}
    \bar{I}_1[y_i,z_i, C, \widehat{C}]
    \\
    \coloneqq 
     -y_i^{(r)*}y_i^{(r)} 
    - i  w y_i^{(r)*} z_i^{(r)} - i w^* z^{(r)*}_iy_i^{(r)}
    \\+ \frac{1}{i} \sum_{n=1}^{\infty}\sum_{a=1}^n 
    \biggl[
        \frac{\kappa_n}{n!} i^n \binom{n}{a}\widehat{C}^a_n(y_i^{(r)*})^a (y_i^{(r)})^{n-a}
        +
        \\
        C_n^a (z_i^{(r)})^a (z_i^{(r)*})^{n-a}
    \biggr]
    .
\end{multline}

Although \siki{e:DefI1} and \siki{e:DefI1bar} are a little bit complicated, we do not have to calculate any more.
In the large-$N$ limit, the saddle point approximation for the integration over $C_n^a$ and $\widehat{C}_n^a$ becomes exact.
From \siki{e:DefI1} and \siki{e:DefI1bar}, it is obvious that the saddle point depends $P_J$ only through $\kappa_n$ i.e. $K(q)$, and so does $\Psi$.

The above result means that the eigenvalue spectrum $\rho$ can be classified by $K(q)$ in the large-$N$ limit.
In other words, we have proved that each class of the eigenvalue spectrum of the large random matrix $J$ has one-to-one correspondence to each universality class.
In particular, when we apply this discussion to the Gauss class, we reproduce the Girko's circular law~\cite{Sommer1988,Girko1985}.

For demonstrations of the above correspondence, we have performed numerical calculations of eigenvalue spectrum for several universality classes, shown in \zu{f:EigenvalueSpectrum}.
As for the Gaussian distribution (\zu{f:EigenvalueSpectrum} (a)), the eigenvalues are located almost uniformly in the unit disk, except the real axis and the region around $z=1.5$.
This uniformness is the result of Girko's circular law.
The dense region of eigenvalues on the real axis is caused by the finite size effect, which has been discussed in Ref.\cite{Sommer1988}.
Furthermore, the dense region around $z=1.5$ is caused from the effect of non-zero mean $(E[J_{ij}]=J_0/N=1.5/N)$ of the random matrix $J_{ij}$.
This is because the peak of this region appears around the mean $\bar{\mu}=J_0$ of $K$, when we change the mean to $J_0=1,1.5,2$.

On the other hand, the eigenvalues of the random matrix sampled from the Gamma distribution (\zu{f:EigenvalueSpectrum} (b)) and symmetrized-Gamma distribution (\zu{f:EigenvalueSpectrum} (c)) do not show such uniformness and are located almost on the origin.
Especially in the former case, the distribution also has a dense region on the real axis and a small peak around $z=1.5$.
From the comparison between (b) and (c), we understand that these characteristics can be regarded as the same effect as in the Gaussian case.
In particular, it is noteworthy that the concentration of eigenvalues at the origin is far from the circular law. 
These differences between the Gauss and Gamma classes are expected to distinguish their response characteristics, such as the dynamic range against input signals.

These numerical results support the correspondence between universality and the eigenvalue spectrum in the large-$N$ limit.

\section{Comments on Practical Choices of Network Structure}
\label{s:Generalization}

In a practical implementation of RCs, we often select the random coupling constant $J_{ij}$ to be not fully connected.
Moreover, it is also customary to choose the spectral radius of $J_{ij}$ as a control parameter, rather than the parameters of the probability distribution.
In this section, we comment on the relation between the theory discussed above and these practical choices of $J_{ij}$.

\subsection{Sparse networks}
\label{s:Sparse}
In \Sec{s:Universality}, the way of parameter tuning was demonstrated only for ``fully connected" networks. 
Whereas previous studies often use a not fully connected network, such as the Erd\H{o}s-R\'enyi network, in the implementation of RC to reduce the computational cost~\cite{Lu2017, Pathak2018}.
In this case, we need to take a different way of tuning from the previous one.
Here, we explain that not fully connected networks also have universality.

To classify the sparsity of the networks, we introduce the average number of connected nodes for every node as $k(N)$ ($0< k(N) \leq N$).
The network is fully connected if $k(N)=N$, dense if $k(N)=O(N)$, and sparse if $k(N)=o(N)$ as $N\to \infty$.
Especially if $k(N)=O(N^0)$, the network is called truly sparse.

In the following, we show that dense and (not truly) sparse networks have universality.
Because each of the coupling constants $J_{ij}$ are set to be zero with the probability of $1-k(N)/N$, 
the probability distribution function of $J_{ij}$ is described as 
\begin{align}
    P_J(J_{ij}) = \qty(1 - \frac{k(N)}{N}) \delta(J_{ij}) + \frac{k(N)}{N} P_s(J_{ij}),
\end{align}
where $P_s$ is some probability distribution function.
Then $K_J(q)$ is given by
\begin{align}
    K_J(q) = \log(1 - \frac{k(N)}{N}  + \frac{k(N)}{N} \exp(K_s)),
    \label{e:KJsparse}
\end{align}
where $K_s(q)$ is the second cumulant generating function of $P_s$. 
Recall that we should tune parameters in the way to satisfy $\lim_{N\to\infty} K_J(q)=0$.
This fact and \siki{e:KJsparse} deduce that $K_s(q)$ should be tuned to satisfy
\begin{align}
    \lim_{N\to \infty} \frac{k(N)}{N} (\exp(K_s(q))-1) = 0.
\end{align}
Therefore, $K_J(q)$ is expressed up to the leading-order in the $N\to\infty$ limit as
\begin{align}
    K_J(q) = \frac{k(N)}{N}\qty(\exp(K_s(q))-1) + (\text{higher order terms}).
\end{align}
With this expression, $K(q)$ is given by
\begin{align}
    K(q) & = \lim_{N\to \infty} k(N)(\exp(K_s(q))-1).
\end{align}

In the dense and (not truly) sparse network, $k(q)$ has an asymptotic form such as
\begin{align}
    k(N) \sim c N^\alpha
\end{align}
with some positive constant and $\alpha\, (0<\alpha \leq 1)$.
Note that the case $\alpha=0$ corresponds to the truly sparse network 
and that the fully connected network is realized in the case $\alpha=c=1$.
We take $k(N)=c N^\alpha$ in the following for simplicity.
Our argument holds straightforwardly for any function $k(N)$ which diverges in the $N\to\infty$, such as $\log N$, 
because all we need to do is exchange $N^\alpha$ to $k(N)$.

Since $k(N)$ diverges as $N\to\infty$, $K_s(q)$ should becomes zero in this limit to keep $K(q)$ finite.
Then, to calculate $K(q)$, we may just consider the limit
\begin{align}
    K(q) & = \lim_{N\to \infty} k(N)K_s(q) = c\lim_{N\to \infty} N^\alpha K_s(q).
    \label{e:KSparse}
\end{align}
This expression is the same as the definition of $K(q)$ (\siki{e:DefK}), where $N$ is replaced to $N^{\alpha}$.
Therefore, the discussion in \Sec{s:LargeNbehaviourK} can be applied for dense or sparse networks, and it is sufficient to determine the $N$-dependence of parameters $\za_a(N)$ as
\begin{align}
    \za_a(N) = (F_{\kappa}^{-1}(\ifrac{\bar{\kappa}_b}{N^\alpha}))_a,
    \label{e:TuningSparse}
\end{align}
where $F^{-1}_\kappa$ is given in \siki{e:NDepPara}.
As a result, \siki{e:DefK} and \siki{e:NDepPara} become \siki{e:KSparse} and \siki{e:TuningSparse} by replacing $N$ to $N^\alpha$, respectively.
Therefore, we find that $K(q)$ in dense or sparse networks can be classified into the same universality classes for the fully connected networks.

\subsection{Universality in controlling spectral radius}
Here we study the relationship between rescaling of the coupling matrix $J_{ij}$ and the universality.
In the previous studies, the matrix $J_{ij}$ is usually rescaled to control its spectral radius, defined as the absolute maximum of its eigenvalues. 
We denote the spectral radius of the matrix $J_{ij}$ by $\rho(J_{ij})$.
The reason for adjusting $\rho(J_{ij})$ is that it plays a crucial role in determining the dynamics and computational performance of the RC.
Actually, in the case $\rho(J_{ij})< 1$, the network holds the echo-state property \cite{Jaeger2001}, which is believed to be necessary for the RC to work well at the learning tasks.

Let us control the spectral radius of $J_{ij}$ in the following way.
We first sample $J'_{ij}$ from some probability distribution $P'(J'_{ij})$ with $N$-independent parameters.
Then, we determine $J_{ij}$ as 
\begin{align}
\label{e:DetermineJ}
J_{ij} = J'_{ij} \times \frac{r}{\rho(J'_{ij})}
\end{align}
to set its spectral radius to be an arbitrary non-negative number $r$, that is, $\rho(J_{ij}) =r$.
This is because $\rho(J_{ij}) = \rho(J'_{ij} \times \frac{r}{\rho(J'_{ij})}) = \frac{r}{\rho(J'_{ij})} \times \rho(J'_{ij}) = r$.

Let us consider the case that the spectral radius scales as
\begin{align}
\label{e:scalingRho}
    \rho(J'_{ij}) \sim c \sqrt{N}
\end{align}
with some positive constant $c$, which is determined from $P'_{ij}$ as $N\to\infty$.
A typical example is the circular law in the random matrix theory \cite{Girko1985,tao2008random}.
It states that an eigenvalue distribution function of a $N\times N$ matrix sampled from i.i.d with zero mean and finite variance is asymptotically given by the probability density function of the uniform distribution on a disk of radius $\sigma \sqrt N$ such as 
\begin{empheq}[left={P_{\mathrm{sc}}(\epsilon)=\empheqlbrace}]{align}
      &\frac{1}{\pi\sigma^2 N} &&\qty(\abs{\epsilon}< \sigma \sqrt{N}),\\
      &\ \ \ \ 0  &&(\text{otherwise}),
\end{empheq}
in the $N\to \infty$ limit.
$\sigma$ is the variance of the sampling probability distribution for the components.
It follows from this law that the spectral radius $\rho$ is $\sigma \sqrt{N}$, the value at the edge of support of $P_{\mathrm{sc}}$.

Given this behavior, we can discuss the effect of rescaling of $J_{ij}$ and the probability distribution of $J_{ij}$.
\siki{e:DetermineJ} and \siki{e:scalingRho} lead to scaling of $J_{ij}$ as $J_{ij} = \frac{r}{c\sqrt{N}} J'_{ij}$.
Let $K'(q)$ be the second cumulant generating function of $J'_{ij}$, then $K_J(q)$ is given by
\begin{align}
    K_J(q) = K'\qty(\frac{r}{c\sqrt{N}} q).
\end{align}
Therefore we get
\begin{align}
    K(q) = \lim_{N\to \infty} NK'\qty(\frac{r}{c\sqrt{N}} q),
\end{align}
which is essentially the same limit as one considered in the (generalized) central limit theorem (\siki{e:CentralLim}).
As a result, we find that $K(q)$ belongs to just the Gauss or stable class when we control the spectral radius of the random coupling matrix $J_{ij}$.

\section{Numerical Calculation}
\label{s:NumCalc}
\begin{figure*}[t]
  \centering
  \begin{tabular}{lll}
(a)  & (b) & (c) \\
    \includegraphics[keepaspectratio, height=0.27\linewidth]
      {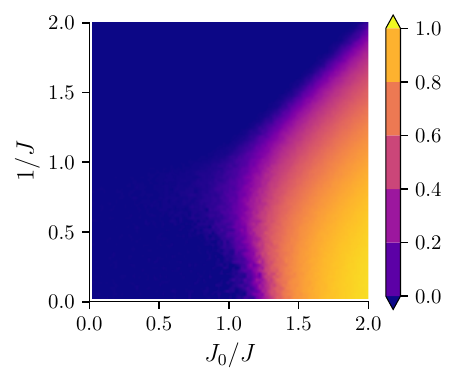} &
    \includegraphics[keepaspectratio, height=0.27\linewidth]
      {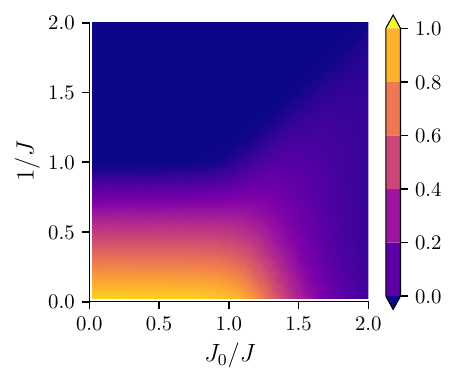} & 
   \includegraphics[keepaspectratio, height=0.27\linewidth]
      {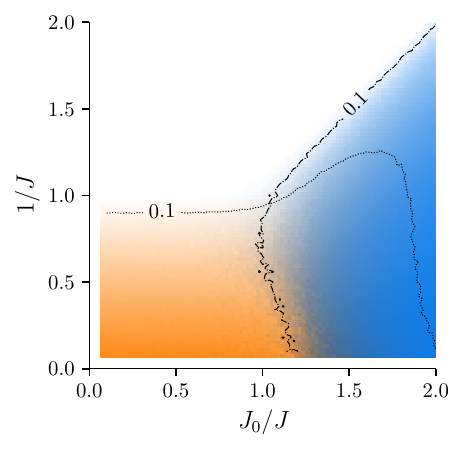}
    \\
(d)  & (e) & (f)  \\
    \includegraphics[keepaspectratio, height=0.27\linewidth]
      {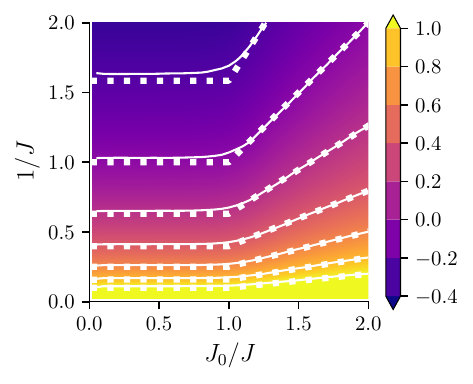} 
      &
     \includegraphics[keepaspectratio, height=0.27\linewidth]
      {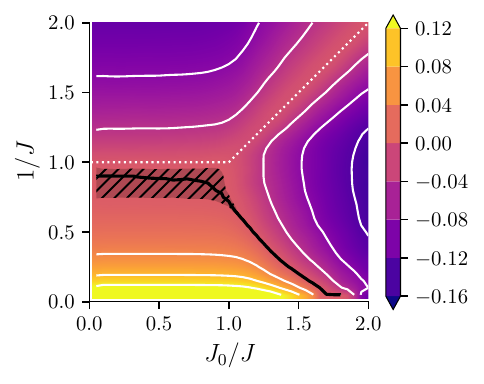} 
      &
      \includegraphics[keepaspectratio, height=0.27\linewidth]
      {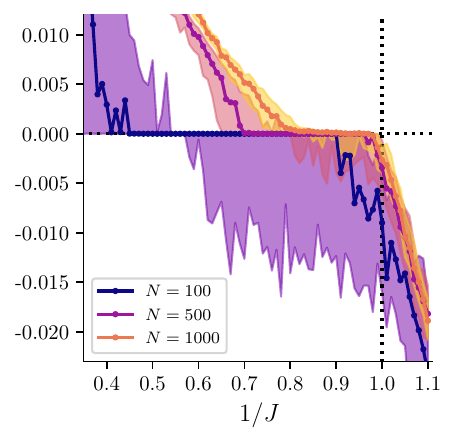}  
  \end{tabular}
 \caption{Phase diagram of the Gaussian network with the network size $N=500$ and the mean $E[J_{ij}]=J_0/N$ and the variance $V[J_{ij}]=J^2/N$.
 (a) Site-mean and (b) Site-variance of  $\phi(r_i)$, which are averaged over 100 trials. 
 (c) Phase diagram of the steady state of the Gaussian network. We show the site-mean (blue) and the site-variance (orange) of $\phi(r_i)$ in the parameter space.
 We set the RGB color values by (R, G, B) $= (255(1-\tilde{\sigma}^2), 255(1-(\tilde{\sigma}^2+\tilde{\mu})/2), 255(1-\tilde{\mu})))$, where $\tilde{\mu}$ and $\tilde{\sigma}^2$ are the ensemble normalized mean and variance of MSE.
 The maximum and minimum values of the normalized quantities are set to be unity and zero, respectively.
  The contours with the dotted and dash-dotted lines indicate $\tilde{\sigma}^2=0.1$ and $\tilde{\mu}^2=0.1$, respectively.
 (d) Color map of the spectral radius of a random matrix for $N=500$ sampled from the Gaussian distribution. The contour lines are indicated by the common logarithm scale and their levels are displayed as labels of the colorbar. The solid lines are empirical contours and the dotted ones are loci of $(1/J, J_0/J)$ satisfying $1/J = 10^{-0.2k} \times \max\{J_0, J\}$ $(k=-2, -1, \ldots, 5)$.
 (e) Color map of the ensemble median of the Lyapunov exponent for Eq.~(\ref{e:discrete eq}) for the Gaussian Network with $N = 500$. The black line shows the zero contour and the gray area indicates that the exponent is in the range $[0, 0.001]$. The white dotted line indicates the theoretical 0 contour of the spectral radius of $J_{ij}$.
 (f) Median of the Lyapunov exponent for $100$ ensembles. Here, we set $J_0=0$ and plot the results for $N = 100$, $500$ and $1000$ by green, orange and blue dots, respectively. The error bars indicate the interquartile range.}
\label{f:PD Gauss}
\end{figure*}

In the previous sections, we have clarified the relation between the network structure, especially the probability distribution of the coupling constant $P_J(J_{ij})$, and the universality of the network dynamics in the large-$N$ limit.
The delta and the Gauss class have already been investigated in detail in previous studies~\cite{Sompolinsky1988,Rajan2010,Schuechker2018,Crisanti2018}.
Importantly, in these cases, we can describe the network dynamics with a closed set of (self-consistent) equations for the one- and two-point correlation functions of the reservoir states. 
Owing to this simplicity, for these classes, we can analytically determine the phase diagram (introduced later in \Sec{s:PD}) in the parameter space~\cite{Sompolinsky1988,Cabana2013,Schuechker2018, Keup2021} and evaluate various dynamical quantities under driving signals, such as the maximum Lyapunov exponent and the memory curve~\cite{Massar2013,Schuechker2018,Haruna2019}.
On the other hand, for more general universality classes such as the Gamma and the stable class, we need to deal with an infinite number of the self-consistent equations for all cumulants, which would be intractable without a numerical approach in general.

For these reasons, in this section, we perform some numerical calculations for the Gauss and the Gamma classes and compare each result to clarify the role of the higher-order statistics in the network dynamics.
Furthermore, we give a discussion on the relation between these results and the edge of chaos.
More specifically, we give the numerical analyses for the following observations, respectively, for both the Gauss and the Gamma classes: (A) The phase diagram and the Lyapunov exponents under no driving input, (B) Common signal-induced synchronization, and time series inference task, 
(C) time series forecasting task using the closed-loop network systems.

As a result, we will show that each universality class has different phase diagram defined by the asymptotic behavior of the reservoir states,
and the phase boundaries is desired areas to improve the computational performance.
These observations suggest the following strategy for designing the RCs to improve their computational performance: $\rm(\hspace{.18em}i\hspace{.18em})$ Following to Table~\ref{t:UnivClass}, determine the proper scaling of the parameters of the probability distribution $P_J$,
$\rm(\hspace{.18em}ii\hspace{.18em})$ By performing numerical simulations, investigate the phase diagram of asymptotic network dynamics in the parameter space obtained in the first step,
$\rm(\hspace{.18em}iii\hspace{.18em})$ Tuning the parameters near the phase boundary obtained in the second step, and execute the computational tasks.

All numerical results below were obtained in the following way unless otherwise stated.
We simulate the discretized version of Eq.~\eqref{network eq} given by
\begin{multline}
\label{e:discrete eq}
    r_i(t+1)
    \\
    =(1-\alpha) r_i(t) + \alpha \qty(\sum_{j=1}^N J_{ij} \phi(r_j(t)) + b_i(t)),
\end{multline}
where $\alpha$ is the learning rate and set to be $\alpha=0.2$ in this study.
We set the network size as $N=500$ throughout this paper, and we average simulation results over 100 trials, each of which has a different ensemble of the random coupling constant $J_{ij}$ and the initial state of the reservoir $r_i(0)$.
Furthermore, for simplicity, we put the activation function as $\phi(x)=\tanh(x)$ in the following. 
Although this choice may lose the generality of our discussion, it enables us, as shown below, to investigate several fascinating dynamical properties of random networks, including the dynamical transition from a fixed point to chaotic behavior and the edge of chaos in the computational performance. 

The numerical codes used in the following calculations are freely available in \cite{github}.

\begin{figure*}[t]
  \centering
  \begin{tabular}{lll}
(a)  & (b)  & (c)\\
    \includegraphics[keepaspectratio, height=0.27\linewidth]
      {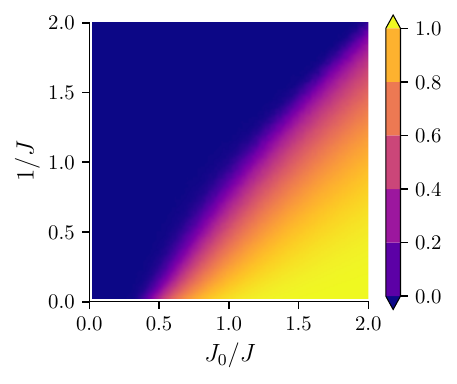} & 
    \includegraphics[keepaspectratio, height=0.27\linewidth]
      {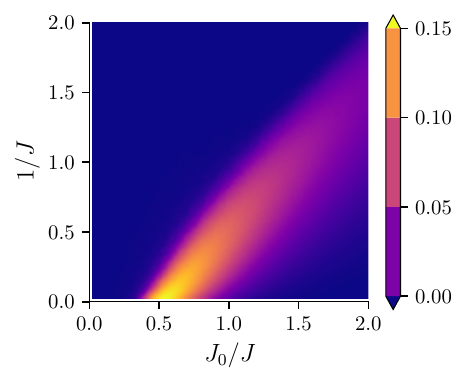} & 
      \includegraphics[keepaspectratio, height=0.27\linewidth]
      {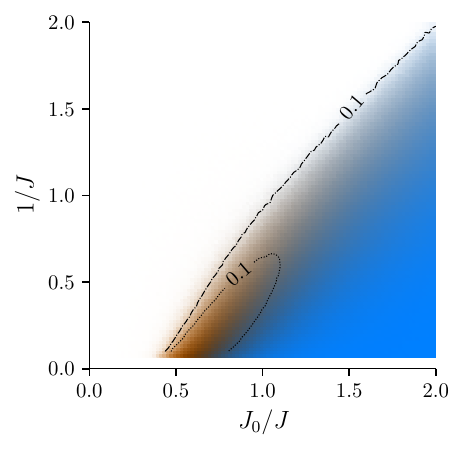} 
       \\
(d)  & (e) & \\
     \includegraphics[keepaspectratio, height=0.27\linewidth]
      {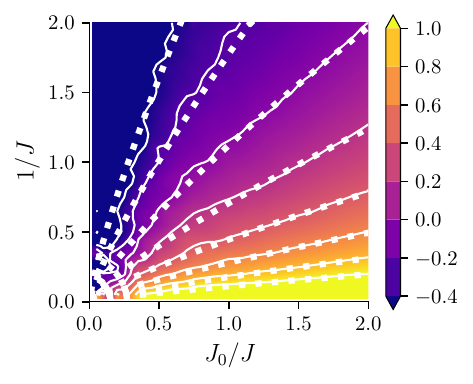} 
      &
     \includegraphics[keepaspectratio, height=0.27\linewidth]
      {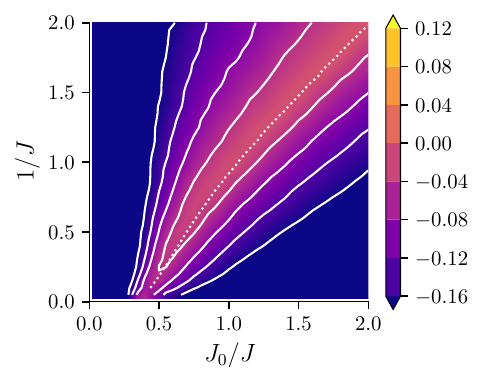}
      &
  \end{tabular}
 \caption{Phase diagram of the Gamma-distributed random neural network with the network size $N=500$ and the mean $E[J_{ij}]=J_0/N$ and the variance $V[J_{ij}]=J^2/N$.
 (a) Site-mean and (b) Site-variance of $\phi(r_i)$, which are averaged over 100 trials.
 (c) Phase diagram of the steady state of the Gamma-distributed random neural network.
 We show the site-mean (orange) and the site-variance (blue) of $\phi(r_i)$ in the parameter space.
 The RGB colors are the same as Fig.~\ref{f:PD Gauss}-(c).
 The contours with the dotted and dash-dotted lines indicate $\tilde{\sigma}^2=0.1$ and $\tilde{\mu}^2=0.1$, respectively.
 (d) Color map of the ensemble mean of the spectral radius of random matrix sampled from the Gamma distribution for $N = 500$. The solid lines are empirical contours and the dotted ones are loci of $(1/J, J_0/J)$ satisfying $1/J = 10^{-0.2k+0.05} \times (J_0/J)^{0.8}$ $(k=-2, -1, \ldots, 5)$.
 (e) The color map of the ensemble median of the Lyapunov exponent for Eq.~(\ref{e:discrete eq}) for the Gamma network with $N=500$. The white dotted line is same as the contour of $\tilde{\mu}^2=0.1$ showed in panel (c).
 }
\label{f:PD Gamma}
\end{figure*}

\begin{figure*}[hbtp]
 \centering
   \begin{tabular}{lll}
(a)  & (b)  & (c)\\
    \includegraphics[keepaspectratio, height=0.27\linewidth]
      {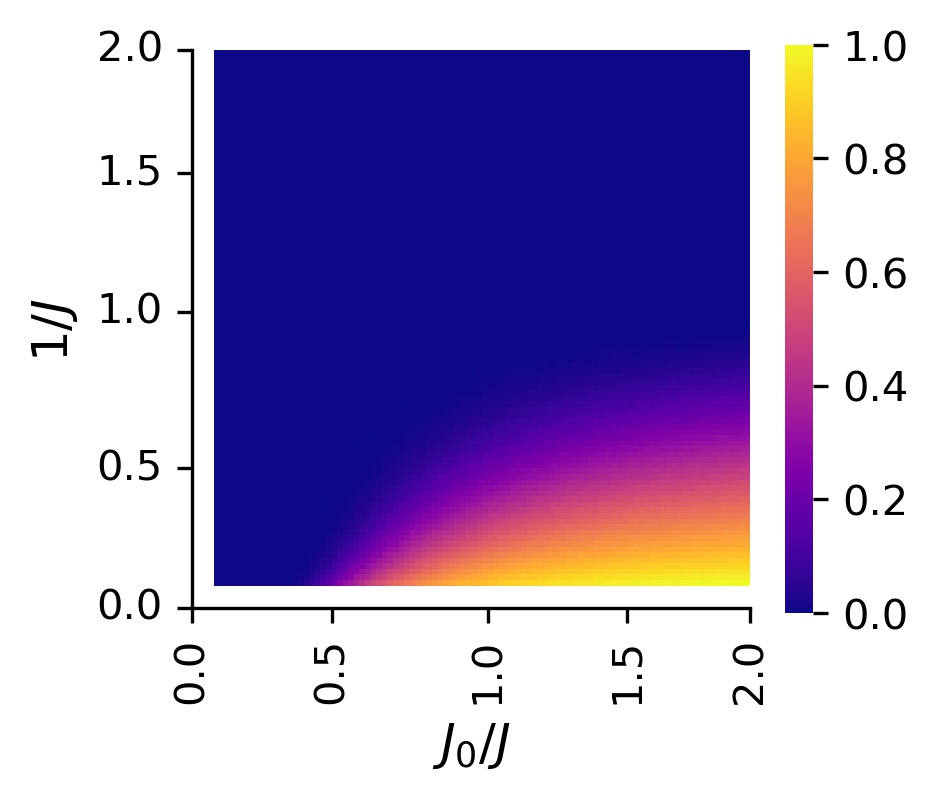} &
    \includegraphics[keepaspectratio, height=0.27\linewidth]
      {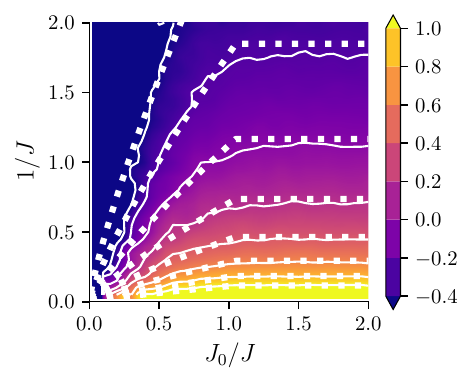}
      &
       \includegraphics[keepaspectratio, height=0.27\linewidth]
      {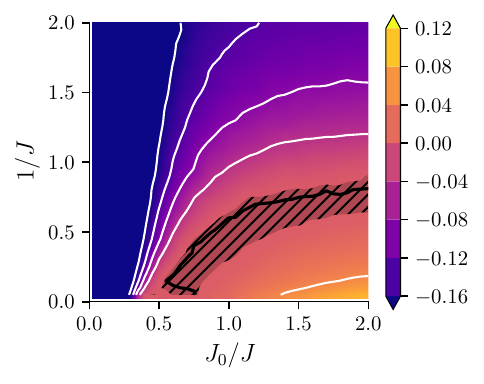}
      \\
  \end{tabular}
 \caption{Phase diagram of the symmetrized Gamma-distributed random neural network with the network size $N=500$ and the mean $E[J_{ij}]=J_0/N$ and the variance $V[J_{ij}]=J^2/N$.
 (a) Phase diagram of the steady state of the symmetrized Gamma-distributed random neural network. We plot site-variance of $\phi(r_i)$, which is averaged over 100 trials.
 (b) color map of the spectral radius of random matrix sampled from the symmetrized Gamma distribution for $N = 500$.
 (c) The color map of the ensemble median of the Lyapunov exponent for Eq.~(\ref{e:discrete eq}) for the symmetrized Gamma network with $N=500$. The gray area indicates that the exponent is in the range $[0, 0.001]$.}
\label{f:PD s-Gamma}
\end{figure*}

\subsection{Phase diagram}
\label{s:PD}
Here we investigate phase diagrams of the steady state of random neural networks under no driving input (we refer to it as merely ``a phase diagram''). 
To this end, for every value in the parameter space, we iterate Eq.~\eqref{e:discrete eq} over enough time steps (5000 steps) for the reservoir state to reach the steady state and record the reservoir state at the final time step $t_f=5000$.
The results for the Gaussian and Gamma networks are shown respectively in Fig.~\ref{f:PD Gauss} and Fig.~\ref{f:PD Gamma}.
Their vertical and horizontal axes are set to be $1/J$ and $J_0/J$ with the mean $E[J_{ij}]=J_0/N$ and the variance $V[J_{ij}]=J^2/N$ for each network.
More specifically, we set $\mu=J_0/N$ and $\sigma^2=J^2/N$ for the Gaussian network and $\theta=J^2/J_0$ and $k=J_0^2/J^2N$ for the Gamma network. 
In panels (a) and (b) in these figures, respectively, we show the site mean and variance of $\phi(r_i(t_f))$, averaged over all trials. 
These figures are combined into a single color map in (c), with the mean and variance of the sites represented in blue and orange, respectively.

We have also confirmed that the phase diagrams discussed here can be reproduced robustly and universally for various system sizes ranging from $N=100$ to $N=1000$, and also for any distribution within an identical class (for example, the Gaussian distribution, uniform distribution, and the Laplace distribution for the Gauss class).
In this sense, the theory of random network dynamics in the large-$N$ limit, discussed in the previous section, gives a good description of their phase diagrams even for large- but finite-size networks.

Figures.~\ref{f:PD Gauss}~(c) and \ref{f:PD Gamma}~(c) can be understood from a physical viewpoint as follows.
The white regime has vanishing site-mean and variance of the reservoir states.
It illustrates the realization of a trivial quiescent regime, where all the network nodes converge to a stable fixed point $r_i=0$ for $i=1,\ldots,N$.
The blue regime is also a quiescent regime, but it has two stable fixed points $r_i=\pm r_*$ $(i=1,\cdots,N)$, where $r_*$ is determined by the parameter values $(1/J,J/J_0)$.
Our simulations sampled the initial reservoir states $r_i(0)$ from the uniform distribution on the interval of $[0,1]$ to make only one fixed point ($r_i=r_*$) appear in the long time limit.
The transition from the white regime to the blue one is analogous to a paramagnetic-ferromagnetic transition in spin systems, and so we refer to each regime as the {\it unpolarized} and the {\it polarized ordered regime}.
On the other hand, the orange (gray) regime, which has zero (non-zero) mean and finite variance of the reservoir states, is more complicated to understand because each regime has different attractor structures, respectively, for the Gauss and Gamma classes.
In the case of the Gauss class, the finite variance suggests the appearance of chaotic dynamics.
It is well-known that the analytical results for $J_{ij}$-averaged networks show the direct transition from the quiescent regime to the chaotic one at the boundary of the white and orange regimes on the contour line of $1/J = 1$ in the large-$N$ limit~\cite{Sompolinsky1988,Crisanti2018}.
However, the actual simulation indicates multiple transitions from a fixed point $r_i=0$ to a limit cycle (torus) or chaotic behavior around $1/J=1$ due to the finite size effect. This observation is qualitatively guaranteed by the following discussion.

To evaluate whether the network dynamics is chaotic or not, we evaluate the Lyapunov exponent of our autonomous model. The Lyapunov exponent quantitatively measures the rate of separation in the time evolution of orbits with different initial states. 
Fig.~\ref{f:PD Gauss}~(e) shows the color map of the median of the Lyapunov exponent of the Gaussian network with $E[J_{ij}] = J_0/N$ and $V[J_{ij}]=J^2/N$ in the $(J_0/J, 1/J)$ space with $N=500$. The black line shows the zero contour and the gray area indicates that the exponent is in the range $[0, 0.001]$.
The tendency of the distribution of the exponent in the parameter space is consistent with the phase diagram Fig.~\ref{f:PD Gauss}~(c).

The random network becomes chaotic in the region below the black line in the sense of the Lyapunov exponent. 
This curve is different from the contour of $J$'s spectral radius of 1 that is represented by the white dotted line in the Fig.~\ref{f:PD Gauss}~(e). 
In the area $1/J<1$, the zero contour of the Lyapunov exponent asymptotically reaches the dotted line, as shown in Fig.~\ref{f:PD Gauss}~(f). Fig.~\ref{f:PD Gauss}~(f) shows the median of the Lyapunov exponents for 100 ensembles for the networks with $J_0=0$, corresponding to the left-end in the two-dimensional phase space, for $N=100$, $500$, $1000$ with green, orange and blue dots, respectively. The error bars represent the interquartile range (IQR). In each case of the network size $N$, the ranges only appear in either the positive or negative regions of the exponent. Thus, one can immediately identify whether the system is ordered or chaotic. The switching point of the sign of the region containing the IQR approaches zero as $N$ increases.

Furthermore, it is noteworthy that Fig.~\ref{f:PD Gauss}~(c) has the similar structure as the phase diagram of the Sherrington-Kirkpatrick (SK) model for spin glass~\cite{Sherrington1975}.
These structures of the phase diagram and analogy with the SK model have been discussed in detail in several previous works~\cite{Sompolinsky1988,Cabana2013,Schuechker2018,Zhang2021}.

Contrary to this, for the Gamma class, higher-order statistics drastically change the attractor structures in the phase diagram.
We obtain a finite variance of the reservoir states in a narrow region in the phase space (Fig.~\ref{f:PD Gamma}~(b)). However, it never indicates the appearance of dynamical states such as a chaotic dynamics or a limit cycle because the Lyapunov exponent of the Gamma networks is always negative in any parameter (Fig.~\ref{f:PD Gamma}~(e)). The absence of chaos is a notable feature of the Gamma network.

In this case, reservoir states converge to a site-dependent fixed point throughout the regime with finite variance in Fig.~\ref{f:PD Gamma}~(c).
Each reservoir state $\{r_i\}$ has a different value at this fixed point, depending on the site and the trial of $J_{ij}$. For this reason, we refer to such a phase as {\it random fixed point} phase. 

Furthermore, while the spectral radius of the connectivity matrix $J_{ij}$ affects the location of the phase boundary for the Gaussian network, such a correspondence does not seem to exist for the Gamma network (see panel (d) in Fig.~\ref{f:PD Gauss} and Fig.~\ref{f:PD Gamma}). Instead, the boundaries' location appears along the contour of the Lyapunov exponent showed in (Fig.~\ref{f:PD Gamma}~(e)).

As another example of networks with the contribution of higher-order statistics, let us consider the symmetrized Gamma network, a network with the connectivity matrix $J$ sampled from the symmetrized Gamma distribution parameterized as the mean $E[J_{ij}]=0$, the variance $V[J_{ij}]=J^2/N$ and the fourth cumulant  $\ev{J_{ij}^4}_c=J^6/(J_0N)$. 
This distribution can be implemented by symmetrizing the Gamma distribution with the mean $E[J_{ij}]=J_0/N$ and the variance $V[J_{ij}]=J^2/N$. 
The components of $J$ in this network distribute symmetrically around the origin and the symmetrized Gamma distribution has non-negligible higher-order cumulants in $K(q)$. 
We show the numerical results of the phase diagram in the Fig.~\ref{f:PD s-Gamma}, where the vertical and horizontal axes are set to be $1/J$ and $J_0/J$.
Fig.~\ref{f:PD s-Gamma}~(c) shows the ensemble median of the Lyapunov exponent with $N=500$ for 100 ensembles. The black line and the gray area are the same as in Fig.~\ref{f:PD Gauss}, respectively.

The symmetrized Gamma network always shows the zero mean of $\phi(r_i)$ for any parameter value and is possible to have the non-zero variance for $J_0/J\gtrsim 0.5$.
We notice that the tendency of the distribution of the exponent is consistent with the phase diagram shown Fig.~\ref{f:PD s-Gamma}~(a) and also with the spectral radius Fig.~\ref{f:PD s-Gamma}~(b). Unlike the Gamma network, we can find the chaotic phase in the bottom right in the phase space. Especially when we fix the value of $J_0/J$ above $0.5$, the phase diagram for $1/J$ has a similar structure to that for the Gaussian network with vanishing mean.
Actually, as shown in Fig.~\ref{f:PD s-Gamma}, numerical results of the maximum Lyapunov exponent suggest that the symmetrized Gamma class shows multiple transitions from an ordered phase to a limit cycle (torus) phase to a chaotic phase, as well as the Gaussian network.

\subsection{Open-loop networks}
In this subsection, we consider what changes arise in the response properties to external signals due to the higher-order statistics and the resulting difference in the phase diagram of each universality class. 
First, we simulate the common signal-induced synchronization under chaotic signals, which is essential for the RC. 
Second, we demonstrate the time series inference task for the chaotic inputs with the RC scheme and show that the best computational performance is achieved near the boundary of the regime with finite site variance.
Throughout this subsection, we use a time series of the x-coordinate $x(t)$ of the Lorenz system 
\begin{empheq}[left=\empheqlbrace]{align}
\displaystyle \dv{x}{t} {(t)} &= \sigma(y(t)-x(t)), \\
\displaystyle \dv{y}{t} {(t)} &= -y(t) + x(t)(\rho-z(t)), \\
\displaystyle \dv{z}{t} {(t)} &= x(t)y(t) - \beta z(t),
\end{empheq}
with the parameters $\sigma=10$, $\rho=28$ and $\beta=8/3$ introduced in~\cite{Lorenz1963} as a chaotic input signal and put $b_i(t)$ in Eq.~\eqref{e:discrete eq} as $b_i(t)=W_{i}x(t)$, where $W_i$ are sampled from independent uniform distribution over $[-1, 1]$.
$x(t)$ are normalized to have zero mean and a unit of variance over the total time steps.

\subsubsection{Common-signal-induced synchronization}
\label{s:CSIS}
Here we analyze the relationship between the phase diagram and the common-signal-induced synchronization.
For this purpose, we iterate Eq.~\eqref{e:discrete eq} under chaotic inputs over 100,000 time-steps and record the reservoir state at the final time step $t_f$.
We take 100 trials of this calculation, more precisely, ten samples of $J$ and ten ensembles of the initial states of $r$ for each $J$.
The initial states are sampled independently from the uniform distribution on the interval $[0,1]$.
Then, we calculate the variance of $\phi(r_i(t_f))$ over all the ensemble of initial states $r_i(0)$ for each $J$, and average it over all sites.
In particular, vanishing variance suggests that common signal-induced synchronization is accomplished in the random network.

\begin{figure}[hbtp]
 \centering
 \includegraphics[keepaspectratio, width=0.8\linewidth]
      {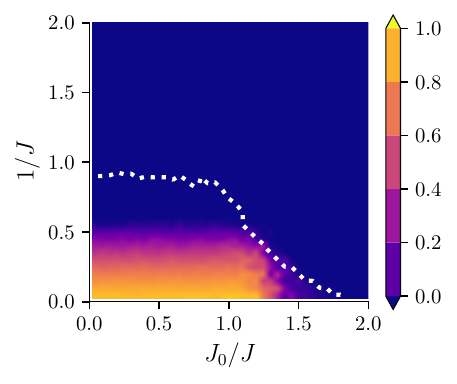}
 \caption{Common signal-induced synchronization for input signal of the Lorenz system in Gaussian network ($N=500$). The color map shows the site averege and $J$-average of the variance of $\tanh(r_i(t_f))$ over all the ensemble of initial state $r_i(0)$. Common signal-induced synchronization is realized only in the regime where the values become zero. 
 The white dotted line represents the zero contour of the Lyapunov exponent of the Gaussian network without input shown in Fig. ~\ref{f:PD Gauss}~(e) for comparison.
 }
\label{f:common signal}
\end{figure}

The numerical result for the Gaussian network is shown in Fig.~\ref{f:common signal}.
Comparing with Fig.~\ref{f:PD Gauss} (e), we readily find that the chaotic phase shrinks by the driving input, especially in the left half in the phase space, which means that the external input effectively decreases the Lyapunov exponent of the network system.
This behavior is known as stimulus-dependent suppression of chaos and investigated analytically, especially for the Gaussian networks in previous studies~\cite{Molgedey1992,Rajan2010,Schuechker2018}.

On the other hand, we can confirm that the Gamma networks show common signal-induced synchronization for any parameter, and thus the ensemble variance of $r_i(t_f)$ becomes zero all over the parameter space.
(Figure for this case is not shown.)
Namely, the input time series always causes common signal-induced synchronization for any parameters in the Gamma networks.

\subsubsection{time series inference task}
\label{s:TSPT}

\begin{figure*}[hbtp]
 \centering
   \begin{tabular}{lll}
(a)  & (b) & (c) \\
    \includegraphics[keepaspectratio, height=0.27\linewidth]
      {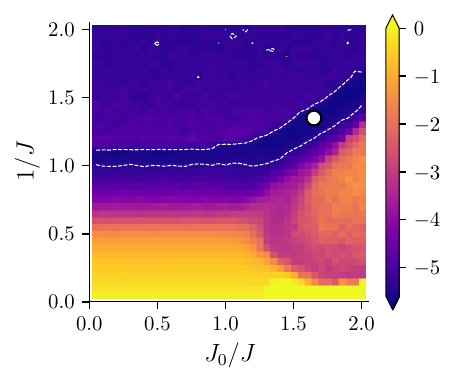} %MSEtest_median_OLP_Nres0500_GaussianM_tanhr_J0J_Linear.pdf
      &
    \includegraphics[keepaspectratio, height=0.27\linewidth]
      {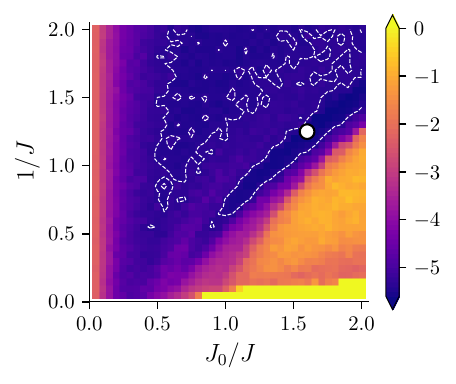} %MSEtest_median_OLP_Nres0500_GammaM_tanhr_J0J_Linear
      &
      \includegraphics[keepaspectratio, height=0.27\linewidth]
      {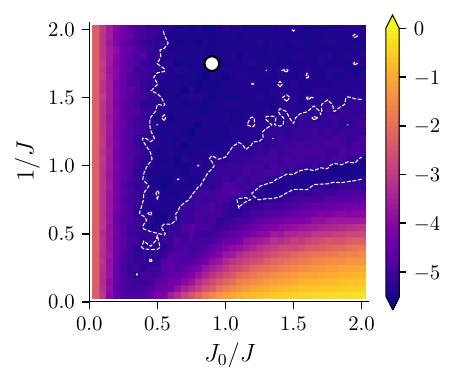} %MSEtest_median_OLP_Nres0500_sGammaM_tanhr_J0J_Linear
      \\
(d)  & (e) & (f)  \\
    \includegraphics[keepaspectratio, height=0.27\linewidth]
      {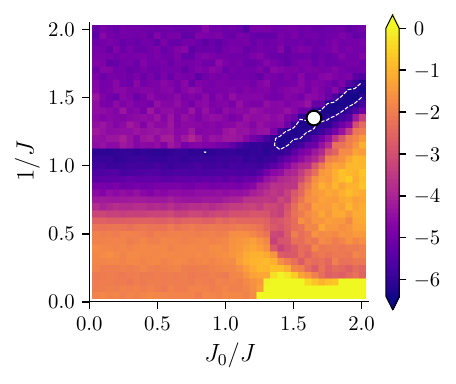} %MSEtest_IQR_OLP_Nres0500_GaussianM_tanhr_J0J_Linear
      &
    \includegraphics[keepaspectratio, height=0.27\linewidth]
      {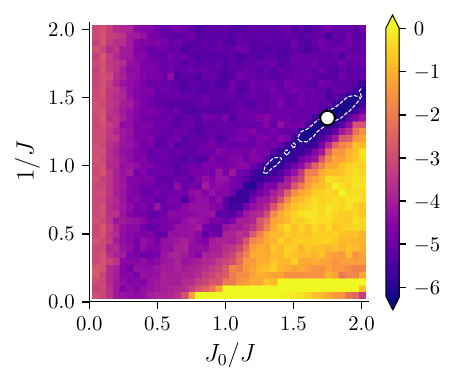} %MSEtest_IQR_OLP_Nres0500_GammaM_tanhr_J0J_Linear
      &
        \includegraphics[keepaspectratio, height=0.27\linewidth]
      {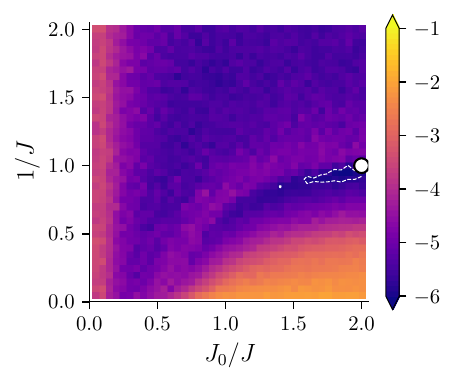} %MSEtest_IQR_OLP_Nres0500_sGammaM_tanhr_J0J_Linear
      \\
      % Visualisation_imshow_MSE_SoA_OLP_Lorenz63_kudpc_J0J_Linear.ipynp
  \end{tabular}
 \caption{Ensemble median (upper) and interquartile range (lower) of the MSE of the time series inference task  for the test data using the Gaussian (a, d), the Gamma (b, e) and the symmetrized Gamma networks (c, f) with $N=500$. 
 The white point and the white dashed line in each figure indicate the minimum point of the MSE and the contour of 95\% level of the minimum, respectively.
 The MSEs plotted here are used in a common logarithm scale.
 }
\label{f:MSE}
\end{figure*}

In this part, we perform a time series inference task for the Lorenz system time series~\cite{Lu2017,Zimmermann2018}.
Our computational scheme is essentially based on that in Ref.~\cite{Lu2017}. 
In the present task, as in Sec.~\ref{s:CSIS}, we use the normalized $x$-coordinate of the Lorenz systems $x(t)$ as input data, and attempt to predict the concurrent $y$-coordinate of the system $y(t)$. 
First, in the training process, we sample a random initial state $r_i(0)$ from the uniform distribution on $[0,1]$ and iterate Eq.~\eqref{e:discrete eq} under the driving signal $x(t)$ up to the transient time $T_{\rm trans}$.
Here we choose $T_{\rm trans}$ to be large enough to guarantee that the reservoir state loses the memory of its initial state by common signal-induced synchronization.
After that, we record all reservoir states generated from Eq.~\eqref{e:discrete eq} over 100,000 $(\eqqcolon T_{\rm train}$) time-steps.
Then we train linear readout parameters $W^{\rm out}_i$ to minimize the mean squared error (MSE) between the estimated values $\hat{y}(t) =\sum_j W^{\rm out}_{j} \phi(r_j(t))$ and the actual value $y(t)$.
Since this is a simple linear regression problem, we can solve it analytically and obtain the optimal value of $W^{\rm out}_i$ as 
\begin{equation}
    W^{\rm out}_{i} =\sum_{j=1}^N (R^{-1})_{ij} B_{j},
\end{equation}
where
\begin{align}
\label{e:learn-Y}
    R_{ij} \coloneqq \sum_{t=T_{\rm tran}}^{T_{\rm train}} \phi(r_i(t)) \phi(r_j(t)), \ 
    B_{j} \coloneqq \sum_{t=T_{\rm tran}}^{T_{\rm train}} \phi(r_j(t)) y(t).
\end{align}

If the training is successfully done, the readout of the reservoir output $\hat{y}(t) =\sum_j W^{\rm out}_{j} \phi(r_j(t))$ should yield a good approximation for the actual value of $y(t)$.
Preparing another time series of the Lorenz system generated from the other random initial state as the test data, we perform the time series inference task and evaluate the MSE.
Here we drop the early time series up to the transient time $T_{\rm trans}$, as in the training case.
We perform this procedure for any parameter values $(1/J,J_0/J)$ over $100$ ensembles of the random coupling constant $J_{ij}$, and lastly, calculate the median and the IQR of MSE over all the ensembles.

We show the numerical results in Fig.~\ref{f:MSE} for the Gauss, Gamma, and symmetrized Gamma classes.
First, paying attention to the results for the Gauss class, we readily notice the remarkable improvement of the computational performance -- a sharp decrease of the median and the IQR of MSE -- around the boundary of the chaotic phase, namely, {\it edge of chaos}~\cite{Bertschinger2004,Bertschinger2005, Legenstein2007a,Legenstein2007b,Busing2010,Toyoizumi2011,Schuechker2018}.
In these previous works other than Ref.~\cite{Bertschinger2005}, they focused mainly on the case where the coupling constant sampled from the Gaussian distribution with vanishing mean.
Thus, it was still unclear what role the non-zero mean value of the coupling plays in computation at the edge of chaos.
As we can understand by comparison with Fig.~\ref{f:PD Gauss} (e) and Fig.~\ref{f:common signal}, the drastic behavior mentioned above appears only in the particular regime where the maximum Lyapunov exponent is near zero and the common signal-induced synchronization is realized. 
We can confirm that similar behavior can be observed even in the symmetrized Gamma class, as shown in Fig.~\ref{f:MSE} (c) and (f).

Further remarkable is that the best computational performance is achieved around the boundary between the polarized ordered and chaotic regimes.
It seems that the minimum of the MSE is attained near the tricritical point $(J_0/J,1/J)=(1,1)$, which is a point at which three-phase coexistence terminates in the phase diagram (Fig.~\ref{f:PD Gauss}(c)).
Interestingly, we get better computational performance in the ordered phases than in the chaotic phase, around the minimum point.
This behavior contrasts with the results on the line $J_0=0$, where the edge of chaos has been most investigated, and the computational performance is relatively better in the chaotic regime (see also Ref.~\cite{Toyoizumi2011,Schuechker2018}).

On the other hand, the Gamma class (Fig.~\ref{f:MSE} (b) and (e)) provides us with a further notable feature.
Even though the Gamma network does not exhibit chaotic dynamics as shown in Sec.~\ref{s:PD}, drastic improvement of the computational performance appears around the boundary of the random fixed point phase.
This behavior is analogous to that at the edge of chaos in the Gaussian network, and the best value of MSE is quantitatively comparable to that in the Gaussian network.
These results suggest that a chaotic phase (or the boundary) is not necessarily needed to improve computational performance. 

In addition, although the fact that the Gamma network holds common-signal-induced synchronization for any parameters, the network has worse inference performance in the bottom-right region in the parameter space (Fig.~\ref{f:MSE}~(b, e).
Thus, the echo-state property is a necessary but not sufficient condition to learn the appropriate readout $W_{\rm out}$.
This fact has yet to be found in the previous studies examining cases for $J_0=0$.

\subsection{Closed-loop networks}
In this subsection, we investigate the performance in a time series forecasting task using the RC with output feedback. 
The procedure is given as follows. 
First, we set a one-step ahead prediction task
\begin{align}
    &
        \begin{multlined}
            r_i(t+1)=
            \\
            (1-\alpha) r_i(t) + \alpha \qty(\sum_{j=1}^N J_{ij} \phi(r_j(t)) + W_i x(t)), \ 
        \end{multlined}
    \label{f:Eq-Reservoir}
    \\
    & \hat x(t+1) = \sum_{j=1}^N W^{\rm out}_{j} \phi(r_j(t+1)),
    \label{f:Eq-Reservoir-readout}
\end{align}
with the input $x(t)$, and we learn $W^{\rm out}_{j}$ in the same way described in Sec.~\ref{s:TSPT}. 
Then, substituting the predicted input $\hat{x}(t)$ for $x(t)$  in Eq.~\eqref{f:Eq-Reservoir}, we obtain the closed-loop system
\begin{multline}
    r_i(t+1)=
    \\
    (1-\alpha) r_i(t) + \alpha \qty(\sum_{j=1}^N \qty(J_{ij} + W_i W^{\rm out}_{j} ) \phi(r_j(t)) ).
    \label{f:Eq-ClosedLoop}
\end{multline}

When starting a forecasting task, we feed the correct input time series $x(t)$ to Eq.~\eqref{f:Eq-Reservoir} for a warm-up until $T_{\rm trans}$ to adjust the reservoir states to the appropriate states.
After that, we stop to provide the proper input.
Then, we run the closed-loop system Eq.~\eqref{f:Eq-ClosedLoop} for $t > T_{\rm trans}$ without the actual input $x(t)$, and we forecast $x(t)$ by $\hat x(t)$ from Eq.~\eqref{f:Eq-Reservoir-readout}.

In this study, we apply this procedure to the normalized $x$-coordinate of the Lorenz system and perform a time series forecasting task.
Since its dynamics is chaotic, the predicted time series is also highly sensitive to the initial conditions.
Therefore, we attempt to forecast the time series until the Lyapunov time defined by $T_{\rm L} = 1/\lambda_{\rm Lorenz}$, where $\lambda_{\rm Lorenz} \approx 0.905$ is the Lyapunov exponent of the Lorenz system.
Then we evaluate the computational performance from the MSE of the test data for $T_{\rm trans} \le t \le T_{\rm trans} + T_{\rm L}$. 

Moreover, since the closed-loop system constructed in the forecasting task is autonomous, it is also important to evaluate the similarity between the RC and the original Lorenz system as a dynamical system.
Ref.~\cite{Pathak2017} suggested the importance of matching their Lyapunov exponents for prediction tasks.
Therefore, we measure the reproducibility of the Lyapunov exponent of the closed-loop system by comparing it to $\lambda_{\rm Lorenz}$ so that we can estimate the degree of the reconstruction of the original dynamics.

\begin{figure*}[hbtp]
 \centering
   \begin{tabular}{lll}
(a) & (b) & (c) \\
    \includegraphics[keepaspectratio, height=0.27\linewidth]
      {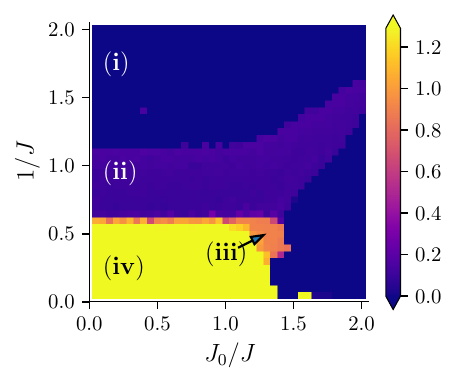} 
      % Visualisation_imshow_LyapunovExp_SoA_CLP_Lorenz63_kudpc_J0J_Linear.ipynb
      &
    \includegraphics[keepaspectratio, height=0.27\linewidth]
      {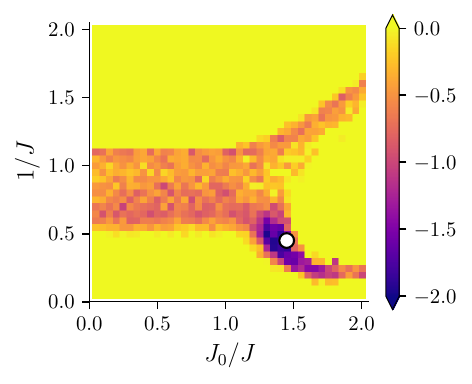} &
    \includegraphics[keepaspectratio, height=0.27\linewidth]
      {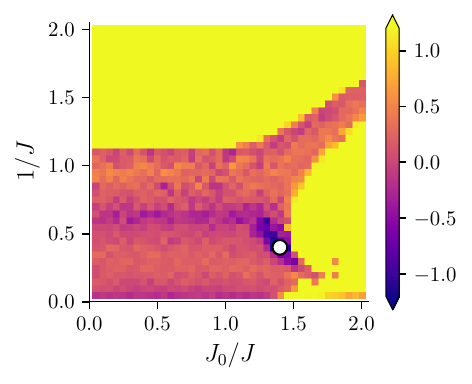}
      \\
(d) & (e) & (f) \\
    \includegraphics[keepaspectratio, height=0.27\linewidth]
      {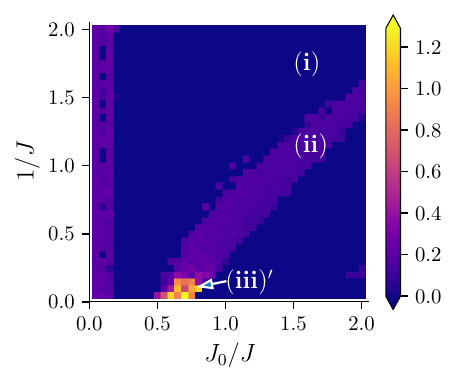} 
      % Visualisation_imshow_LyapunovExp_SoA_CLP_Lorenz63_kudpc_J0J_Linear.ipynb
      &
    \includegraphics[keepaspectratio, height=0.27\linewidth]
      {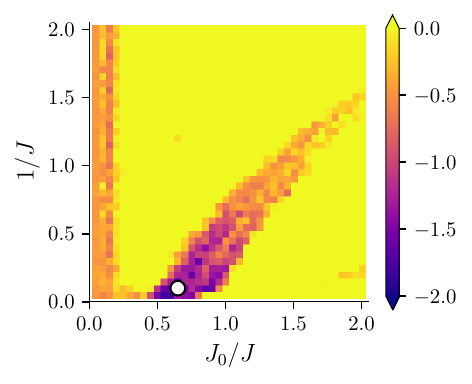} &
    \includegraphics[keepaspectratio, height=0.27\linewidth]
      {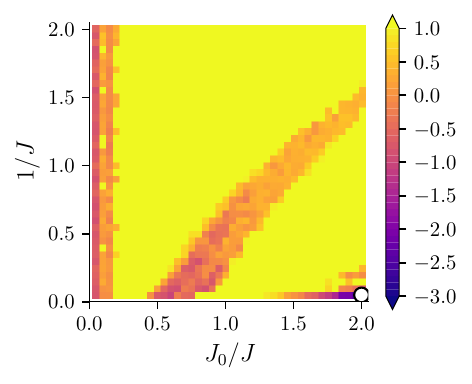}
    \end{tabular}
 \caption{(a, d) Ensemble median of the Lyapunov exponent of the closed loop dynamics, (b, e) Ensemble median and (c, f) interquartile range of the MSE of test data of the closed loop system constructed by (upper) Gaussian and (lower) Gamma networks with $N=500$. The values showed in (b, c, e, f) are given in the common logarithmic scale.}
\label{f:MSE closedloop}
\end{figure*}

In Figs.~\ref{f:MSE closedloop}, we show our results for the Lyapunov exponent and the MSE of our prediction task for each pair of the parameters $(1/J, J_0/J)$ over $100$ ensembles of the random coupling constant $J_{ij}$ and the initial states $r_i$.
These figures include the median (b, e) and the IQR (c, f) of MSE over all the ensembles for the Gaussian and the Gamma networks. 

The ensemble medians of the Lyapunov exponents are shown in Figs.~\ref{f:MSE closedloop}~(a) and (d) for the Gaussian and the Gamma networks, respectively.
The parameter space can be divided into four areas, numbered in ascending order of the ensemble medians from (i) to (iv) in Fig.~\ref{f:MSE closedloop}~(a) for the Gaussian case, and into three areas, numbered from (i), (ii) and (iii)' in Fig.~\ref{f:MSE closedloop}~(d) for the Gamma case.

The darker blue area (i), found at a distance from the phase boundaries in both unpolarized and polarized ordered regimes, shows the negative exponents; namely, the system is not chaotic. 
This non-chaotic property is common to both the Gaussian and the Gamma cases.
Thus, the forecasting system in this parameter area is unsuitable for emulating the original chaotic dynamics.
We can also find severely large errors in the area (i) even in figures for the Gaussian (Figs.~\ref{f:MSE closedloop}~(b, c)) and the Gamma networks (Figs.~\ref{f:MSE closedloop}~(e, f)).
Therefore, adopting parameter pairs $(J_0, J)$ from this area is not suitable for our prediction task.

The purple area (ii) is close to the 1-contour of the spectral radius of the connecting matrices shown in Fig.~\ref{f:PD Gauss}~(d) and Fig.\ref{f:PD Gamma}~(d), the closed-loop networks have positive Lyapunov exponents.
However, they are considerably smaller than that of the Lorenz system $\lambda_{\rm Lorenz}$.
Consequently, the network can only predict with the accuracy of MSE of $10^{-1}$.

The orange area (iii), found near the tricritical point in the Gaussian case, achieved the best performance in reproducing the Lyapunov exponent.
The network with the parameter $(J_0, J)$ from this area also accomplishes the smallest ensemble median and IQR of MSE of the test data. 

Similar to this result, the best reproducibility in the Gamma case is obtained in the area (iii)', which is near the phase boundary between the random fixed point phase and the the other phases in Fig.~\ref{f:PD Gamma}.
Remind that the Gamma network does not have a chaotic regime, unlike the Gaussian case. However, the optimal parameter can still be found near the phase boundary. 
The smallest IQR of MSE is found in the bottom right in the parameter space (Fig.~\ref{f:PD Gamma}~(f)).
All the reservoir states take values close to $1$ due to the magnitude of $J_{ij}$ being considerably large in this parameter area.
Therefore, it would be more appropriate to take the parameters from the area (iii)' for the computational performance.

The yellow area (iv), particular to the Gaussian case, found in the chaotic regime, indicates that the closed-loop system has a significantly greater Lyapunov exponent than $\lambda_{\rm Lorenz}$.
The systems with the parameters from this area are likely to have a larger Lyapunov dimension \cite{KaplanYorke} than the original one.
Thus they hardly generate a time series qualitatively similar to that produced by the original attractor.

As observed in previous task for  the open-loop system, the above results for the closed-loop system also imply that RCs show their best performance around the phase boundary in both cases of the Gauss class and the Gamma class.

\section{Discussion and Conclusion}
\label{s:Conclusion}

In this paper, we have described a path integral approach to
the dynamics of general randomly connected neural networks beyond the Gaussian assumption.
We have provided a dynamical mean-field method to these problems, which describes the network dynamics exactly in the large-$N$ limit and classified these networks into several universality classes.
Under this approximation, the inter-nodes coupling term in Eq.~\eqref{network eq} is replaced by an effective random noise $\eta(t)$, as described in Eq.~\eqref{e:effective EOM}, and its time-correlation functions are determined by the $N$-leading contribution of the cumulant generating function $K(q)$ for $J_{ij}$.
At the end of Sec.~\ref{s:Universality}, we have also shown that this function $K(q)$ is closely related to the eigenvalue spectrum of the random coupling matrix $J_{ij}$ in the large-$N$ limit.
In particular, throughout this paper, we have emphasized that higher-order statistics play non-negligible roles in the dynamics of general neural networks, such as the Gamma and the stable class.
This fact contrasts that the usual Gaussian networks are described only by the one- and two-point correlation functions of the reservoir states.
In the last part of Sec.~\ref{s:Universality}, we have proved that each universality class has a one-to-one correspondence with the eigenvalue spectrum of the random coupling matrix. 
In \Sec{s:Generalization}, we have studied the universality of not fully connected networks and showed that they could be classified with the same universality classes of fully connected networks.
We have also argued that networks with the controlled spectral radius belong to the Gauss class.

In the latter half of this paper, we have performed several numerical simulations to demonstrate how the contribution of higher-order statistics changes the dynamical or informational properties.
First, we have confirmed that the Gamma (and symmetrized Gamma) network has different attractor structures in the parameter space compared with the Gaussian network.
We have also mentioned a finite size effect in the attractor structures of the Gaussian network with an analysis of the maximum Lyapunov exponents.
We have confirmed no direct transition of the attractor from the fixed point to the chaotic behavior in a finite Gaussian network, but a limit-cycle phase appears between them.
Interestingly, the differences in the attractor structures have appeared clearly in the computational performance, and especially, the best performance has been realized in the boundary of the phase with finite variance in Fig.~\ref{f:PD Gauss} and Fig.~\ref{f:PD Gamma}. 
In particular, these observations in the Gamma network suggest that the boundary of a chaotic phase is not necessarily needed to improve the computational performance, and instead, the complexity of the attractor structure is crucial.
These results seem to be beyond our conventional understanding and may provide a new perspective on the discussion of the edge of chaos.

Finally, we would like to give our future perspective.
As a major premise, it is crucial to understand the origin of the dynamical or informational functionality of structured networks, such as brains or trained artificial neural networks, in the research of neurophysics or machine learning.
In these networks, essentially, we cannot neglect the influence of higher-order statistics, and their effects are never tractable in the picture based on the Gaussian network model.
We believe that our work might provide some clues to understand their role in network dynamics.
It is interesting to investigate the relation between the eigenvalue spectrum of $J_{ij}$ and the dynamical range of random networks.
It is also curious to determine in which types of tasks higher-order statistics play an essential role.
These issues will be an interesting future work bridging the random network models and more realistic structured network systems.

Another important direction is to develop a method to calculate the measure of computational performance, such as the mean squared error in the path integral framework.
Primarily, a recent study~\cite{Belkin2019} has reported that the double decent phenomenon occurs when the node size $N$ becomes extremely large, i.e., the computational performance first gets worse and then gets better.
This phenomenon may be understood from the point of view of large-$N$ expansion in the path integral framework with such a  method.

We have given some examples of the universality classes in \Sec{s:ExamUC}, and the complete classification of them is left as another future problem.
Although this classification can be done with which cumulant does not vanish in principle, it is non-trivial what classes are realized under some constraints such as finiteness of cumulants, the number of parameters, differentiability of the cumulant generating function with respect to the parameters.

Furthermore, it is left to be solved as a purely mathematical problem to determine the concrete form of the eigenvalue spectrum density in the large-$N$ limit for each universality class.

\section*{Acknowledgement}

We are thankful to Taro Toyoizumi and Terufumi Yamaguchi for their valuable discussions. We also thank Hiroshi Kokubu and Yoshitaka Saiki for providing helpful comments from a mathematical viewpoint of chaotic dynamical systems. This work is partly
supported by JSPS KAKENHI (Grants 20J22612, 21J14825, 20K03747 and 22H03553) and MACS program at Kyoto University. Part of the numerical calculation of this study was done on the supercomputer of ACCMS, Kyoto University.

J.H. and R.T. contributed equally to this work.

\appendix
\renewcommand{\theequation}{\Alph{section}.\arabic{equation}}
\section*{Appendix}

\section{Derivation of Path-Integral Representation}
\label{a:DetailDerivation}
We briefly review the path-integral formalism in random networks in this Appendix.
In particular, we show the detailed calculation to derive \siki{e:DMF_generating_fn}.

\subsection{Auxiliary fields}
\label{s:AuxFields}
Supposed that $K(q)$ can be described in the Taylor series expansion for $q$, the action $\bar{S}[r,\hat{r}]$ generally has a series of time-nonlocal terms ($n=1,2,\cdots$)
\begin{align}
    &\frac{1}{N}\sum_j K\qty(\int dt\ \hat{r}_i\phi(r_j))
    \\
    &=
    \frac{1}{N}
    \sum_{n=0}^{\infty}
    \frac{\kappa_n}{n!}i^n\sum_j\left(\int dt\ \hat{r}_i\phi(r_j)\right)^n 
    \\
    &
\begin{multlined}
     =\frac{\kappa_n}{n!}i^n \int dt_1 \cdots \int dt_n\ \hat{r}_i(t_1)\cdots  \hat{r}_i(t_n)
    \\ 
    \times\left(\frac1N \sum_j\phi(r_j(t_1))\cdots\phi(r_j(t_n))\right),
\end{multlined}
\end{align}
where $\kappa_n\coloneqq i^{-n} d^nK/dq^n\eval{}_{q=0}$.
$\kappa_n$ correspond to $N$ times the $n$-th order cumulant of the probability distribution $P_J$. 
These nonlocal terms can be simplified by introducing the auxiliary fields $C_n(t_1,\cdots,t_n)$ and $\widehat{C}_n(t_1,\cdots,t_n)$ with the identity
\begin{align}
    1 &= \int DC_n \delta \left(C_n - \frac1N \sum_j\phi(r_j(t_1))\cdots\phi(r_j(t_n))\right)\\
    &= \frac{N}{2\pi}\int D\widehat{C}_n DC_n \\
    &\ \ \ \ \exp\left[-i \hat{C_n}\left(NC_n -  \sum_j\phi(r_j(t_1))\cdots\phi(r_j(t_n))\right)\right].
\end{align}
Performing these transformations, we obtain the averaged generating function and the effective action in the following form:
\begin{align}
\label{e:path_integral3}
    Z[b,\hat{b}] = \int  \qty(\prod_{n=1}^{\infty} DC_nD\widehat{C}_n )
    e^{iN S_1[C,\widehat{C};b,\hat{b}]}.
\end{align}
$S_1$ is defined as
\begin{multline}
    S_1[C,\widehat{C};b,\hat{b}] \coloneqq -W[C,\widehat{C};b,\hat{b}]\\
    +i \sum_{n=1}^{\infty}  \int dt_1 \cdots \int dt_n C_n(t_1,\cdots,t_n)\widehat{C}_n(t_1,\cdots,t_n),
\end{multline}
and $W$ is given by
\begin{multline}
    NW[C,\widehat{C};b,\hat{b}] \coloneqq \\
    \sum_{i=1}^N \log\qty[ \int D\hat{r}_iDr_i 
    \exp(i\bar{S}_1[r_i,\hat{r}_i, C, \widehat{C}] + \int dt\, \hat{b}_ir_i)],
\end{multline}
where $\bar{S}_1$ is the single node contribution to $W$:
\begin{multline}
\label{e:DefSbar}
    \bar{S}_1[r_i,\hat{r}_i, C, \widehat{C}] \coloneqq 
    \int dt\  \sum_{i=1}^N\hat{r}_i \qty(-\dv{t} r_i - r_i + b_i)
    +
    \\ \frac1i \sum_{i=1}^{N}\sum_{n=1}^{\infty}\left[  \frac{\kappa_n}{n!} i^n \int [dt_j]_n C_n(t_1,\cdots,t_n)\hat{r}_i(t_1)\cdots\hat{r}_i(t_n)+\right.\\
    \left. \int [dt_j]_n
    \widehat{C}_n(t_1,\cdots,t_n)\phi(r_i(t_1))\cdots\phi(r_i(t_n)) \right]
    .
\end{multline}
We denote $\prod_{j=1}^n dt_j$ as $[dt_j]_n$.
It should be noted that the functional $W[C,\widehat{C};b,\hat{b}]$ is the generating function of connected time-correlation functions of $r_i$ and $\hat{r}_i$ weighted by the factor $\exp({i\bar{S}_1[r_i,\hat{r}_i, C, \widehat{C}]+\int dt \hat{b_i}r_i})$.

\subsection{Dynamical Mean-Field Approximation}

In the thermodynamic limit, the path integral with respect to $C$ and $\widehat{C}$ can be evaluated by the saddle point approximation, which is so-called the Dynamical Mean-Field (DMF) approximation~\cite{Sompolinsky1988, Crisanti2018, Schuechker2018}.
The saddle point trajectories $C_n^*(t_1,\cdots,t_n)$ and $\widehat{C}_n^*(t_1,\cdots,t_n)$ of the action $S_1[C,\widehat{C};b,\hat{b}]$ dominate the integral and the generating functional~Eq.~\eqref{e:path_integral3} is approximately evaluated in the following form:
\begin{align}
Z[b,\hat{b}] \sim e^{iNS_1[C^*,\widehat{C}^*;b,\hat{b}]}.
\end{align}
The saddle point trajectories $C_n^*(t_1,\cdots,t_n)$ and $\widehat{C}_n^*(t_1,\cdots,t_n)$ are determined to satisfy the following saddle-point equations:
\begin{align}
\eval{\frac{\delta S_1}{\delta C_n}}_{C=C^*,\widehat{C}=\widehat{C}^*} = \eval{\frac{\delta S_1}{\delta \widehat{C}_n}}_{C=C^*,\widehat{C}=\widehat{C}^*}=0.
\end{align}
These equations lead to
\begin{empheq}[left=\empheqlbrace]{align}
    i\widehat{C}_n^*(t_1,\cdots,t_n) &= \frac{1}{N}\sum_{i=1}^{N}
      \frac{\kappa_n}{n!} \ev{\hat{r}_i(t_1)\cdots
    \hat{r}_i(t_n)}_*\\
    C_n^*(t_1,\cdots,t_n) &= \frac{1}{N}\sum_{i=1}^{N}
      \ev{\phi(r_i(t_1))\cdots\phi(r_i(t_n))}_*,
\end{empheq}
where $\ev{\cdots}_*$ denotes the average over the paths weighted by the factor $e^{i\bar{S}_1[r_i,\hat{r}_i, C^*, \widehat{C}^*] + \int dt \hat{b}_ir_i}$. 
This means that the saddle point of the auxiliary fields $i\widehat{C}^*_n,C^*_n$ are given by the site-average of $n$-point correlation functions of $\hat{r}_i(t)$ or $\phi(r_i(t))$.
Here we readily find that \siki{e:correl_hatr} implies that $\widehat{C}^*_n=0$ is the solution to the self-consistent equations.

Under the DMF approximation, the generating function is separated into single-node contributions, 
$Z[b,\hat{b}]\sim \prod_i Z_i[b_i,\hat{b}_i]$, and each contribution is given as follows: 
\begin{align}
&Z_i[b_i,\hat{b}_i] = \int D\hat{r}_iDr_i \exp(i\bar{S}_* + \int dt \hat{b}_ir_i),
\end{align}
which is \siki{e:DMF_generating_fn}.

\section{Wideness of Gauss Class}
\label{a:wideness}
In this Appendix, we give a theorem related to the wideness of the Gauss Class, which states that if the distributions satisfy some conditions, their $K$ has a Gauss-like form.
\begin{theorem}
If (i) a distribution has finite mean $\mu$ and variance $\sigma^2$,  (ii) $\Nt$ is two or more at some of it zeros, (iii) its second cumulant generating function $K_J$ is first-order differentiable at $\mu=\sigma^2=0$, and (iv) there exists a probability distribution $P_K$ whose has $K$ as the second cumulant generating function, then the form of $K$ at the zero is given by the following form:
\begin{align}
    K = i\bar{\mu}q -\frac{\bar{\sigma}^2}{2}q^2 + \bar{\sigma}^2 R(q),
    \label{e:KGLike}
\end{align}
where $R(q)$ is some function of $q$ and $\sorder{q}$ quantity around $q=0$.
\end{theorem}

In the following, we give the proof.
We denote parameters to tune as $c_n$ for $n=1,2,\ldots,\Nt$.
Because of the supposition (i) and (ii), we can regard the mean $\mu$ and variance $\sigma^2$ as parameters instead of $c_1$ and $c_2$, without loss of generality.

From the supposition (iii) and Taylor's theorem, $K_J$ is expressed as 
\begin{multline}
    K_J(q;\mu,\sigma,\za_{a\geq 3}) = K_J(q;0,0,\za_{a\geq 3}) 
    \\
     + \mu \eval{\pdv{K_J}{\mu}}_{\mu=\sigma^2=0}
     + \sigma^2 \eval{\pdv{K_J}{\sigma^2}}_{\mu=\sigma^2=0}
     + \sorder{\mu,\sigma^2}.
\end{multline}
It is obvious that we should tune $\mu$ and $\sigma^2$ in the same way as \siki{e:GaussAssum}, like
\begin{align}
    \mu = \frac{\bar{\mu}}{N}, \sigma^2 = \frac{\bar{\sigma}^2}{N}.
\end{align}
As for $\za_{a\geq 3}$, we do not have to specify their tuning in the following argument.

Then, $K(q)$ becomes
\begin{align}
K = \bar{\mu} F_\mu(q) + \bar{\sigma}^2 F_\sigma(q) + F(q),
\label{e:KWideGauss}
\end{align}
where 
\begin{align}
F_\mu(q) &\coloneqq {\pdv{K_J}{\mu}}(q;0,0,\za_{a\geq 3}(\infty)),\\
F_\sigma(q) &\coloneqq {\pdv{K_J}{\sigma^2}}(q;0,0,\za_{a\geq 3}(\infty)),\\
F(q) & \coloneqq \lim_{N\to \infty} NK_J(q;0,0,\za_{a\geq 3}(N)).
\end{align}
It should also be noted that $F_\mu, F_\sigma$ and $F$ do not depend on $\bar{\mu}$ and $\bar{\sigma}^2$, and also that $F_\mu(q)$ is expanded around $q=0$ as $F_\mu = iq + \sorder{q^2}$, and $F_\sigma(q)$ is $F_\sigma = -q^2/2 + \sorder{q^2}$ by their definitions.

Under the supposition (iv) 
% \footnote{
% 
% }
, let us consider the limit $\bar{\sigma}^2 \to 0$.
Note that because we get $K(q)$ as a result of the limit of $K_J(q)$, this supposition is not necessarily obvious.
With this limit, $P_K$ satisfies \footnote{
Here, we have assumed that domain of definition of $X$ is $(-\infty,\infty)$, which does not lose generality.
If support of $P_K$ is $[a,b]$, it is enough to take $P_K(X)$ as zero for $X < a$ and $X>b$.
}
\begin{align}
    0 = \bar{\sigma}^2 = \int_{-\infty}^{\infty} dX (X - \bar{\mu})^2 P_K(X) .
\end{align}
This equation is saturated only if $P_K(X) = 0$ for $X \neq \bar{\mu}$, which means that the support of $P_K(X)$ is a point $X=\bar{\mu}$.
Therefore, by the definition of the Dirac delta function, it follows that
\begin{align}
    \eval{P_K(X)}_{\bar{\sigma}^2=0} = \delta(X -\bar{\mu}).
    \label{e:RelPKDelta}
\end{align}
Because the second cumulant generating function of the Dirac delta function is $\exp(i\bar{\mu}q)$, \siki{e:RelPKDelta} states that
\begin{align}
    \eval{K(q)}_{\bar{\sigma}^2=0} = i\bar{\mu}q.
    \label{e:KDeltaLim}
\end{align}
Comparing \siki{e:KDeltaLim} and \siki{e:KWideGauss}, we get a dramatic relation
\begin{align}
    F_\mu(q) = iq, F(q) =0.
\end{align}
Remember that $F_\mu(q)$ and $F(q)$ are {\it independent} of $\bar{\sigma}^2$.

Finally, we find that $K(q)$ has the form of \siki{e:KGLike},
% \begin{align}
%     K = i \bar{\mu} q  - \frac{\bar{\sigma}^2}{2} q^2 + \bar{\sigma}^2 R(q),
%     % \label{e:KGLike}
% \end{align}
where $R(q)$ is defined as $R(q) \coloneqq F_\sigma - (-q^2)/2$, and is $\sorder{q^2}$ quantity around $q=0$.
This is the end of the proof.

Moreover, we give a sufficient condition that $R(q)$ becomes zero, i.e. $K(q)$ belongs to the Gauss Class, although we have not yet identified necessary one for it.
\begin{theorem}
If $P_K$ is expressed as 
\begin{align}
    \eval{P_K(X)}_{\bar{\mu}=0,c_{a\geq 3}=c_{a\geq 3}(\infty)} = \bar{\sigma}^{-1}  f \qty(\frac{X}{\bar{\sigma}})
\end{align}
with some function $f$, then $R(q)=0$.
\end{theorem}

It is very easy to show this theorem.
Under the above condition, $K(q)$ is calculated as 
\begin{align}
    &\eval{K(q)}_{\bar{\mu}=0,c_{a\geq 3}=c_{a\geq 3}(\infty)}  \\
    &\coloneqq \log \qty(\int_{-\infty}^\infty dX  \exp(iqX)P_K(X) )\\
    & = \log \qty(\int_{-\infty}^\infty dX  \bar{\sigma}^{-1}\exp(iqX) f \qty(\frac{X}{\bar{\sigma}}) )\\
    & = \log \qty(\int_{-\infty}^\infty dY  \exp(i\bar{\sigma} q Y) f(Y) )\\
    & = \log \tilde{f}(\bar{\sigma} q).
\end{align}
We have changed the integration variable $X$ to $Y\coloneqq X\sigma^{-1}$ in the third equality, and defined the Fourier transform of $f$ as $\tilde{f}$.
This $\tilde{f}$ is expanded around $q=0$ as 
\begin{align}
    \tilde{f}(\bar{\sigma} q) = 1 - \frac{\bar{\sigma}^2}{2}q^2 + \sorder{(\bar{\sigma} q)^2}.
    \end{align}
We have used the fact that $\tilde{f}$ is the characteristic function of $P_K$, from which $\tilde{f}(0)=1$ follows, and the fact that $\bar{\mu}$ is set to zero, which results in $d\tilde{f}/dq(0)=0$.
Therefore $K(q)$ satisfies
\begin{align}
    \eval{K(q)}_{\bar{\mu}=0,c_{a\geq 3}=c_{a\geq 3}(\infty)}  
    = 1 - \frac{\bar{\sigma}^2}{2}q^2 + \sorder{(\bar{\sigma} q)^2},
    \quad \label{e:KTaylor}
\end{align}
because $\log(1+x) = x + \order{x^2}$ holds around $x=0$.

Comparing the linear term of \siki{e:KTaylor} and \siki{e:KGLike} with respect to $\bar{\sigma}^2$, we get $R(q)=0$.
This is the end of the proof of Theorem 2.

\section{How to tune parameters and Renormalization group}
\label{a:renormalization}
In this Appendix, we give general discussion of parameter-tuning.
This is essentially same as renormalization group analysis in field theories in condensed matter or high-energy physics.

\begin{align}
    N K_J (q;c_a(N)) = K(q) + \sorder{N^{0}}
\end{align}

If we $N\pdv{N}$ and ignore $\sorder{N^{0}}$ quantities, we get 
\begin{align}
    K_J(q) + \sum_a \beta_a \pdv{K_J(q)}{c_a} = 0,
    \label{e:RGeq}
\end{align}
which represents $N$-dependence of  $K_J(q)$, called ``renormalization group equation" in field-theoretical physics.
where $\beta_a$ is defined as
\begin{align}
    \beta_a \coloneqq N\pdv{c_a(N)}{N},
\end{align}
called ``beta function", which represents $N$-dependence of parameters.

If we change parameters via diffeomorphism of 
\begin{align}
    c_a = c_a(c'_a)
\end{align}
then we get
\begin{align}
    \beta'_{a} = N\pdv{c'_a}{N} = \sum_b N\pdv{c_b}{N} \pdv{c'_a}{c_b} 
    = \sum_b \beta_b \pdv{c'_a}{c_b}
\end{align}
and 
\begin{align}
    \pdv{K_J}{c'_a} = \sum_b \pdv{c_b}{c'_a}\pdv{K_J}{c_b}
\end{align}

\begin{align}
    \sum_a \beta'_a \pdv{K_J}{c'_a} = \sum_{a,b,c}\beta_b \pdv{c'_a}{c_b} \pdv{c_c}{c'_a} \pdv{K_J}{c_c}
    = \sum_c \beta_c \pdv{K_J}{c_c} 
\end{align}
we have used 
\begin{align}
\sum_a \pdv{c'_a}{c_b} \pdv{c_c}{c'_a}=\delta_{bc},
\end{align}
which shows the renormalization group equation is invariant under parameter-changing, which can be regarded as diffeomorphism.
% using this invariance, we can assume without loss of generality that $K_J$ is

Requiring that the RG equation should hold for arbitrary $q$, 
let us derive RG equations of cumulants when all of them are finite.
Remember that RG equation comes from the Large-$N$ scaling of $K_J$, which is given by $\order{N^{-1}}$.
Therefore, cumulants $\kappa_n$ should have scaling of $N$ like
\begin{align}
    \kappa_n \coloneqq \frac{i^{-n}}{n!}\eval{\pdv[n]{K_J}{q}}_{q=0} = \order{N^{-1}} \quad \text{or} \quad \sorder{N^{-1}},
\end{align}
because the generating function ($K$) of them is $\order{N^{-1}}$.
This leads to \footnote{
Note that equality in \siki{e:RGeq} holds only after taking the limit of $N=\infty$.
\siki{e:RGn} follows from this relation
\begin{align}
\lim_{N\to\infty} N\pdv{N} \log(c_1 N^\alpha + c_2 N^\beta) =\max (\alpha,\beta)
\end{align}
}
\begin{align}
    N\pdv{\log\abs{\kappa_n}}{N} = \sum_{a=1}^\Nt \beta_a \pdv{\log\abs{\kappa_n}}{c_a} \leq -1
    \label{e:RGn}
\end{align}
These is a simultaneous linear equation of $\beta_a$.
The range of summation can be reduced to $N=1,\ldots,\Nt$ because 
\begin{align}
    \pdv{K_J(q;c_b)}{c_b} = 0 
\end{align}
holds when $K_J(q;c_b)=0$ for arbitrary value of $c_b$, 

We do not have to determine $\beta_a$ exactly if we want to know just large-$N$ behaviour of parameters.
The large-$N$ behaviours of $\beta_a$ can be discussed as follows.
Because $K(c_a(N)=c_a(\infty))$ is zero, $\log(K)$ and therefore its derivative diverge with $\delta c_a=0$.
From this fact, it is expected that
\begin{align}
\pdv{c_a}\log \kappa_n \sim \order{(\delta c_a)^{-1}}.
\label{e:BehaveLogKappa}
\end{align}

Even if \siki{e:BehaveLogKappa} is not $\order{(\delta c_a)^{-1}}$ quantity, we can apply diffeomorphism for new parameters $c'_a$ to satisfy \siki{e:BehaveLogKappa}.
Therefore we assume that \siki{e:BehaveLogKappa} holds without loss of generality.
Precise discussion comes in the following paragraphs.

Singularity in \siki{e:BehaveLogKappa} must be canceled by $\beta_a$ for \siki{e:RGn} to hold, which leads to
\begin{align}
    \beta_a \sim \order{\delta c_a}.
\end{align}
Therefore the leading-order terms in $\beta_a$ are linear about parameters.

Let us evaluate $\beta_a$ concretely.
We define $R_{nb}$ as
\begin{align}
    R_{nb} \coloneqq \pdv{\log \kappa_n}{c_b}
\end{align}
then \siki{e:RGn} is rewritten as
\begin{align}
    \sum_b R_{nb}\beta_b \leq  -1.
    \label{e:Rbeta}
\end{align}
It is important that \siki{e:Rbeta} should hold for $n\in\mathbb{N}$.
We choose some set of $R_{nb}$ which saturate the inequality, and denote their linearly-independent components as $M_{ab}\coloneqq R_{i_a b}$ for $a, i_a\in \mathbb{N}$.
That is, $M_{ab}$ satisfies
\begin{align}
    \sum_b M_{ab}\beta_b =  -1.
    % \label{e:Rbeta}
\end{align}
These are simultaneous linear equations for $\beta_a$.
Because of the linearly-independence of $M_{ab}$, there exists inverse matrix of $M_{ab}$.
Denoting this as $M^{-1}_{ab}$, \siki{e:Rbeta} can be solved by
\begin{align}
    \beta_ a = - \sum_{b=1}^\Nt M^{-1}_{ab}.
\end{align}

Because $M_{ab}$ transforms via diffeomorphism as
\begin{align}
    M_{ab} \to \sum_{d} M_{ad} \pdv{c_{i_{d}}}{c'_{i_b}}
\end{align}
and therefore $M_{ab}^{-1}$ transforms as
\begin{align}
    M_{ab}^{-1} \to \sum_{d} \pdv{c'_{i_a}}{c_{i_{d}}} M_{db}^{-1} ,
\end{align}
we can take parameters to satisfy 
\begin{align}
    M^{-1}_{ab} = \order{\delta c_a}.
\end{align}
In the following, we neglect higher order terms.
Let us define the $\order{\delta c_a}$ terms as 
\begin{align}
M^{-1}_{ab} = \sum_d f_{abd} \delta c_d,
\end{align}
and then $\beta_a$ can be determined as
\begin{align}
    \beta_a = - \sum_b r_{ab} \delta c_b,
\end{align}
where 
\begin{align}
r_{ab}&\coloneqq \sum_c f_{acb}.
% \\ &= -\sum_c \pdv{c_b} R^{-1}_{ac}
% \\ &= \sum_c R^{-1}_{ad} \qty(\pdv{c_b} R_{de}) R^{-1}_{ec}
% \\ &= \sum_c R^{-1}_{ad} \pdv[2]{\log \kappa_d}{c_b}{c_e} R^{-1}_{ec}
\end{align}
Then from the definition of $\beta_a$, parameters should satisfy
\begin{align}
    N\pdv{N} \delta c_a = - \sum_b r_{ab}\delta c_b.
    \label{e:RGeqPara}
\end{align}

\siki{e:RGeqPara} describes large-$N$ behaviour of $\delta c_a$ near $N=\infty$.
Let us solve it and determine them.
We define $\vec{x}_a$ as left eigenvector of $r_{ab}$ with eigenvalue of $\epsilon_a$ , which obeys
\begin{align}
    \sum_b (\vec{x}_{a})_b r_{bc} = \epsilon_a (\vec{x}_{a})_c,
\end{align}
and $\gamma_a$ as 
\begin{align}
    \gamma_a \coloneqq \sum_b (\vec{x}_a)_b \delta c_b.
\end{align}
Large-$N$ of $\gamma_a$ is determined from
\begin{align}
    N\pdv{N} \gamma_a & = -\epsilon_a \gamma_a,
\end{align}
whose solution is given by
\begin{align}
    \gamma_a(N)  = \bar{\gamma}_a N^{-\epsilon_a} 
\end{align}
with some constants $\bar{\gamma}^a$.
Then, letting $\vec{y}_a$ be dual vectors of $\vec{x}^a$, such as right eigenvector of $r_{ab}$, $\delta c_a$ can be calculated by
\begin{align}
    \delta c_a =\sum_{b} (\vec{y}_a)_b \gamma_b.
\end{align}
Therefore, the solution of \siki{e:RGeqPara} is 
\begin{align}
    \delta c_a(N) = \sum_{b} (\vec{y}_a)_b \bar{\gamma}_b N^{-\epsilon_b},
\end{align}
which gives complete $N$-dependence of $\delta c_a$.

Here we give an easy example of the above procedure.
Let us consider uniform distribution, whose probability distribution function is given by
\begin{empheq}[left={P_J(X)=\empheqlbrace}]{align}
\begin{aligned}
     &  \frac{1}{b-a}\quad \text{for}\ a<X<b,
     \\ & 0 \quad \text{otherwise},
\end{aligned} 
\end{empheq}
and whose second cumulant generating function is
\begin{align}
    K_J = i\frac{b+a}{2}q + \log(\frac{\sin \frac{b-a}{2}q}{\frac{b-a}{2}q}).
\end{align}
Its cumulants are given by
\begin{empheq}[left={\kappa_n=\empheqlbrace}]{align}
\begin{aligned}
    &\frac{a+b}{2} \quad (n=1)\\
    & \alpha_n (b-a)^n \quad (n\geq 2)
\end{aligned}
\label{e:UniKappa}
\end{empheq}
with some constants $\alpha_n$.
These satisfy
\begin{align}
    & \pdv{\log \kappa_1}{a}=\pdv{\log \kappa_1}{b}=\frac{1}{b+a}, \\
    & \pdv{\log \kappa_n}{a}=-\pdv{\log \kappa_n}{b}=\frac{n}{b-a} \quad (n\geq 2).
    \label{e:UniDelKappa}
\end{align}
Because zero of $K_J$ is given by $a=b=0$, we should choose two among $\kappa_n$s and impose
\begin{align}
    N\pdv{\log\abs{\kappa_n}}{N} = -1.
\end{align}

Let us discuss which cumulants saturates \siki{e:Rbeta}
If we adopt both from $\kappa_{n\geq 2}$, there is no solution to \siki{e:RGeqPara}.
So we choose $\kappa_1$ and $\kappa_{n\geq 2}$, which leads to
\begin{align}
    N\pdv{N} \log(b+a) = N\pdv{N} \log(b-a)^n = -1.
    \label{e:UniRGn}
\end{align}
There is another constraint for $\kappa_n$.
$\kappa_n$s which does not saturate \siki{e:Rbeta} must obey
\begin{align}
    N\pdv{N}\log \abs{\kappa_{m\neq 1,n}} < -1.
\end{align}
From \siki{e:UniKappa} and \siki{e:UniRGn}, this inequailty can be calculated as
\begin{align}
     N\pdv{N} \log(b-a)^m = -\frac{m}{n} < -1
    \label{e:UniConst}
\end{align}
for $m\in \mathbb{N}\cap m\neq 1,n$.
\siki{e:UniConst} holds if and only if $n=2$.
Therefore, we should choose $\kappa_1$ and $\kappa_2$ to saturate \siki{e:Rbeta}.

According to \siki{e:UniDelKappa}, $M_{ab}$ is given by
\begin{align}
    M_{11} = M_{12} =\frac{1}{b+a},
    M_{21} = -M_{22} = - \frac{2}{b-a},
\end{align}
where we have taken $i_1 = 1$ and $i_2 = 2$.
The inverse matrix of $M_{ab}$ is given by
\begin{align}
    M^{-1} = \frac{1}{4}\mqty( 2(b+a) & -b-a\\ 2(b+a) & b-a)
\end{align}
Note that $M^{-1}$ is linear in terms of $a$ and $b$.
Then $\beta_a$ is given by
\begin{align}
    \mqty(\beta_1 \\ \beta_2)
    = - \frac{1}{4}\mqty(3 & 1 \\ 1 & 3) \mqty(a \\ b),
\end{align}
and $r_{ab}$ is
\begin{align}
    r = \frac{1}{4}\mqty(3 & 1 \\ 1 & 3).
\end{align}
Right eigenvectors of $r_{ab}$ and their eigenvalue are
\begin{align}
    \vec{x}_1 &= \frac{1}{\sqrt{2}}\mqty(-1 \\ 1),\quad \epsilon_1 = \frac{1}{2},\\
    \vec{x}_2 &= \frac{1}{\sqrt{2}}\mqty(1 \\ 1),\quad  \epsilon_2 = 1.
\end{align}
Because $r$ is symmetric matrix, right- and left-eigenvectors are identical ($\vec{x}_a = \vec{y}_a$).
Therefore $N$-dependence of $a$ is given by
\begin{align}
    a = \sum_{b=1}^2 (\vec{y}_1)_b \bar{\gamma}_b N^{-\epsilon_b}
    = -\frac{\bar{\gamma}_1}{\sqrt{2}} N^{-1/2} + \frac{\bar{\gamma}_2}{\sqrt{2}} N^{-1},
\end{align}
and one of $b$ is
\begin{align}
    b  = \sum_{b=1}^2 (\vec{y}_2)_b \bar{\gamma}_b N^{-\epsilon_b}
     = \frac{\bar{\gamma}_1}{\sqrt{2}} N^{-1/2} + \frac{\bar{\gamma}_2}{\sqrt{2}} N^{-1}.
\end{align}
Note that mean ($\mu = \kappa_1$) and variance ($\sigma^2 = \kappa_2$) scales as
\begin{align}
    \mu = \frac{\bar{\gamma}_2}{\sqrt{2}}\frac{1}{N}, 
    \quad 
    \sigma^2 = \frac{\bar{\gamma}_1}{6} \frac{1}{N},
\end{align}
which agree with \siki{e:GaussAssum} with identification of
\begin{align}
    \bar{\mu} = \frac{\bar{\gamma}_2}{\sqrt{2}}, 
    \quad 
    \bar{\sigma}^2 = \frac{\bar{\gamma}_1}{6}.
\end{align}

\newpage
\bibliographystyle{utphys}
\bibliography{citation}

\end{document}